\documentclass[floatfix,footinbib,reprint,twocolumn,prx,aps,a4paper,preprintnumbers,amsmath,amssymb,longbibliography]{revtex4-1}

\usepackage{graphicx}
\usepackage{subfigure}
\usepackage{enumitem}  
\usepackage{bbold} 		
\DeclareGraphicsExtensions{.png,.pdf}
\usepackage{epsfig}
\usepackage{bm}% bold math
\usepackage{color}
\usepackage{makeidx}
\usepackage[colorlinks=true,linkcolor=blue,citecolor=blue,urlcolor=blue]{hyperref}
\usepackage{siunitx}
\usepackage{xcolor}
\usepackage{soul}
\setstcolor{red}
\usepackage{makecell}

\usepackage[bbgreekl]{mathbbol}
\DeclareSymbolFontAlphabet{\amsmathbb}{AMSb}%
\DeclareSymbolFontAlphabet{\mathbb}{AMSb}
\usepackage{amsfonts}
\usepackage{mathbbol}
\usepackage[utf8]{inputenc}
\usepackage[USenglish]{babel}

\begin{document}
\title{\bf \textit{Ab initio} calculation of the magnetic Gibbs free energy of materials using magnetically constrained supercells}

\author{Eduardo Mendive-Tapia{\email[]{mendive@mpie.com, e.emendive.tapia@fz-juelich.de}},$^{1,2}$  Jörg Neugebauer,$^{1}$ and Tilmann Hickel$^{1,3}$}
\affiliation{$^1$Department of Computational Materials Design, Max-Planck-Institut für Eisenforschung, 40237 Düsseldorf, Germany}
\affiliation{$^2$Peter Gr\"{u}nberg Institut and Institute for Advanced Simulation, Forschungszentrum J\"{u}lich \& JARA, D-52425 J\"{u}lich, Germany}
\affiliation{$^3$BAM Federal Institute for Materials Research and Testing, 12489 Berlin, Germany}

\begin{abstract}
We present a first-principles approach for the computation of the magnetic Gibbs free energy of materials using magnetically constrained supercell calculations.
Our approach is based on an adiabatic approximation of slowly varying local moment orientations, the so-called finite-temperature disordered local moment picture.
It describes magnetic phase transitions and how electronic and/or magnetostructural mechanisms generate a discontinuous (first-order) character.                      
We demonstrate that the statistical mechanics of the local moment orientations can be described by an affordable number of supercell calculations containing noncollinear magnetic configurations.
The applicability of our approach is illustrated by firstly studying the ferromagnetic state in bcc Fe. We then investigate the temperature-dependent properties of a triangular antiferromagnetic state stabilizing in two antiperovskite systems Mn$_3$AN (A = Ga, Ni). Our calculations provide the negative volume expansion of these materials as well as the \textit{ab initio} origin of the discontinuous character of the phase transitions, electronic and/or magnetostructural, in good agreement with experiment.
\end{abstract}
%%%%%%%%%%%%%%%%%%%%%%%%%%%%%%%%%%%%%%%%%%%%%%%%%%%%%%%%%%%%%%%%%%%%%%%%%%%%%%%%%%%

\maketitle

%%%%%%%%%%%%%%%%%%%%%%%%%%%%%%%%%%%%%%%%%%%%%%%%%%%%%%%%%%%
%%%%%%%%%%%%%%%%%%%%%%%%%%%%%%%%%%%%%%%%%%%%%%%%%%%%%%%%%%%
\section{Introduction}
\label{Intro}
%%%%%%%%%%%%%%%%%%%%%%%%%%%%%%%%%%%%%%%%%%%%%%%%%%%%%%%%%%%
%%%%%%%%%%%%%%%%%%%%%%%%%%%%%%%%%%%%%%%%%%%%%%%%%%%%%%%%%%%

Solid-state magnetic materials form the basis of many current and upcoming technologies, and are thus subject to a field of research that is in continuous development. Some examples are caloric refrigeration~\cite{PhysRevLett.78.4494,MoyaNatMat2014,PhysRevB.95.104424} and gas liquefaction~\cite{doi:10.1063/5.0006281}, permanent magnets in engines and other electronic equipment, and spintronic technology including magnetic field sensors, neuromorphic computing, and different devices for storage and processing of information~\cite{PhysRevLett.61.2472,Fert2017,Zhang_2020,Song2020,AFMJung,Wadley587}. The central component for their functionality is the formation and exploitation of complex magnetic structures, such as  topologically protected skyrmions~\cite{Muhlbauer915,FeCoSiYu}, antiferromagnetic~\cite{PhysRevLett.113.157201} and helimagnetic~\cite{Zhang2017,Mackintosh1} states, and high coercivity in ferromagnets~\cite{doi:10.1063/1.333572,doi:10.1063/1.333571} and ferrimagnets~\cite{PhysRevMaterials.1.024411,PhysRevLett.120.097202}. The fundamental understanding and prediction of this broad range of magnetic states is, therefore, crucial to advance and create new functional mechanisms.

The properties of these magnetic states can strongly depend on temperature. In fact, sometimes their stabilization is linked to the presence of thermal fluctuations~\cite{Muhlbauer915,PhysRevB.103.024410} and their functionality usually involves magnetic phase transitions driven by temperature changes.
The development of a first-principles tool describing temperature-dependent magnetic states and phase transitions among them is a challenging task but also a necessity to reliably predict and explain magnetic phenomena in materials. Major work has been done for the description of thermal fluctuations of local magnetic moments, including both their transverse~\cite{0305-4608-15-6-018,PhysRevB.89.054427,PhysRevB.74.144411} and longitudinal degrees of freedom~\cite{PhysRevB.95.054426,PhysRevB.102.014402}. Many of these theoretical developments also account for a finite-temperature magnetoelastic or magneto-phonon coupling~\cite{PhysRevB.85.125104,PhysRevB.102.144101,PhysRevLett.121.125902,npjTanaka2020}.
However, most of them focus on the paramagnetic state, i.e.\ a high-temperature limit in which the local magnetic moment orientations are fully disordered away from a description of intermediate temperatures.

In this work we present a computational approach to calculate the \textit{ab initio} Gibbs free energy of a magnetic material,
%%%%%%%%%%%%%%%%%%%%%%%%%%%%%%%%%%%%%%%%%%%%%%%%%%%%%%%%%%%
%\begin{widetext}
\begin{equation}
%\begin{split}
    \mathcal{G}_\text{tot}
%   = 
%   \Bigg\langle
%   k_\text{B}T\ln \left[P_\text{tr}(\{\hat{\textbf{e}}_n\})\right]
%   +E_\text{int}(\{\hat{\textbf{e}}_n\},V)
% -\textbf{H}\cdot\sum_n \mu_n\hat{\textbf{e}}_n
% \Bigg\rangle_\text{tr}
%    +pV
=
%&
U_\text{mag}
%(\varepsilon_{\alpha\beta})
-TS_\text{mag}
-\sum_n\textbf{H}\cdot\big\langle \mu_n\hat{\textbf{e}}_n
 \big\rangle
 +\sigma_{\alpha\beta}\varepsilon_{\alpha\beta},
 %\\
%= &
%\Big\langle E_\text{int}(\{\hat{\textbf{e}}_n\},\varepsilon_{\alpha\beta})\Big\rangle_\text{tr}
%TS_\text{mag} \\
%& -\sum_n\textbf{H}\cdot\big\langle \mu_n\hat{\textbf{e}}_n
% \big\rangle_\text{tr}
% +\sigma_{\alpha\beta}\varepsilon_{\alpha\beta}
 %%   +pV,
 %%   \text{ with }
 %%   S_\text{mag}=\langle\ln \left[P_\text{tr}(\{\hat{\textbf{e}}_n\})\right]\rangle_\text{tr}
%\end{split}
\label{Eq_GIntro}
\end{equation}
%\end{widetext}
%%%%%%%%%%%%%%%%%%%%%%%%%%%%%%%%%%%%%%%%%%%%%%%%%%%%%%%%%%%
whose minimization provides the dependence on temperature of the most stable magnetic states for given values of an external magnetic field $\textbf{H}$ and an applied mechanical stress $\sigma_{\alpha\beta}$, the latter causing a material deformation described by a strain tensor $\varepsilon_{\alpha\beta}$.
$\{\hat{\textbf{e}}_n\}$ are unit vectors prescribing the orientations of the local magnetic moments, located at lattice sites $\{n\}$ and with magnitudes $\{\mu_n\}$.
The central tenet of our approach is to assume that $\{\hat{\textbf{e}}_n\}$ vary very slowly in comparison with $\{\mu_n\}$ and with the underlying interacting electron system.
This means that an \textit{ab initio} magnetic energy can be specified for different configurations $\{\hat{\textbf{e}}_n\}$ and that $\mathcal{G}_\text{tot}$ can be obtained by carrying out ensemble averages over $\{\hat{\textbf{e}}_n\}$, denoted by $\langle\dots\rangle$.
For example, the internal magnetic energy is
%%%%%%%%%%%%%%%%%%%%%%%%%%%%%%%%%%%%%%%%%%%%%%%%%%%%%%%%%%%
\begin{equation}
U_\text{mag}
%(\varepsilon_{\alpha\beta})
 = \Big\langle E_\text{int}(\{\hat{\textbf{e}}_n\}
%,\varepsilon_{\alpha\beta}
)
\Big\rangle 
 =\int\prod_n[d\hat{\textbf{e}}_n]
P(\{\hat{\textbf{e}}_n\})
E_\text{int}(\{\hat{\textbf{e}}_n\}
%,\varepsilon_{\alpha\beta}
),
\label{Eq_UmagIntro}
\end{equation}
%%%%%%%%%%%%%%%%%%%%%%%%%%%%%%%%%%%%%%%%%%%%%%%%%%%%%%%%%%%
where
$P(\{\hat{\textbf{e}}_n\})\propto\exp[-\beta E_\text{int}]$
is a Boltzmann probability distribution for the local moment orientations. An associated magnetic entropy forms the second term in the right hand side of Eq.\ (\ref{Eq_GIntro}) and is given by $S_\text{mag}=-k_\text{B}\langle\ln \left[P(\{\hat{\textbf{e}}_n\})\right]\rangle$, $k_\text{B}$ being the Boltzmann constant.

\begin{figure}[t]
\centering
\includegraphics[clip,scale=0.22]{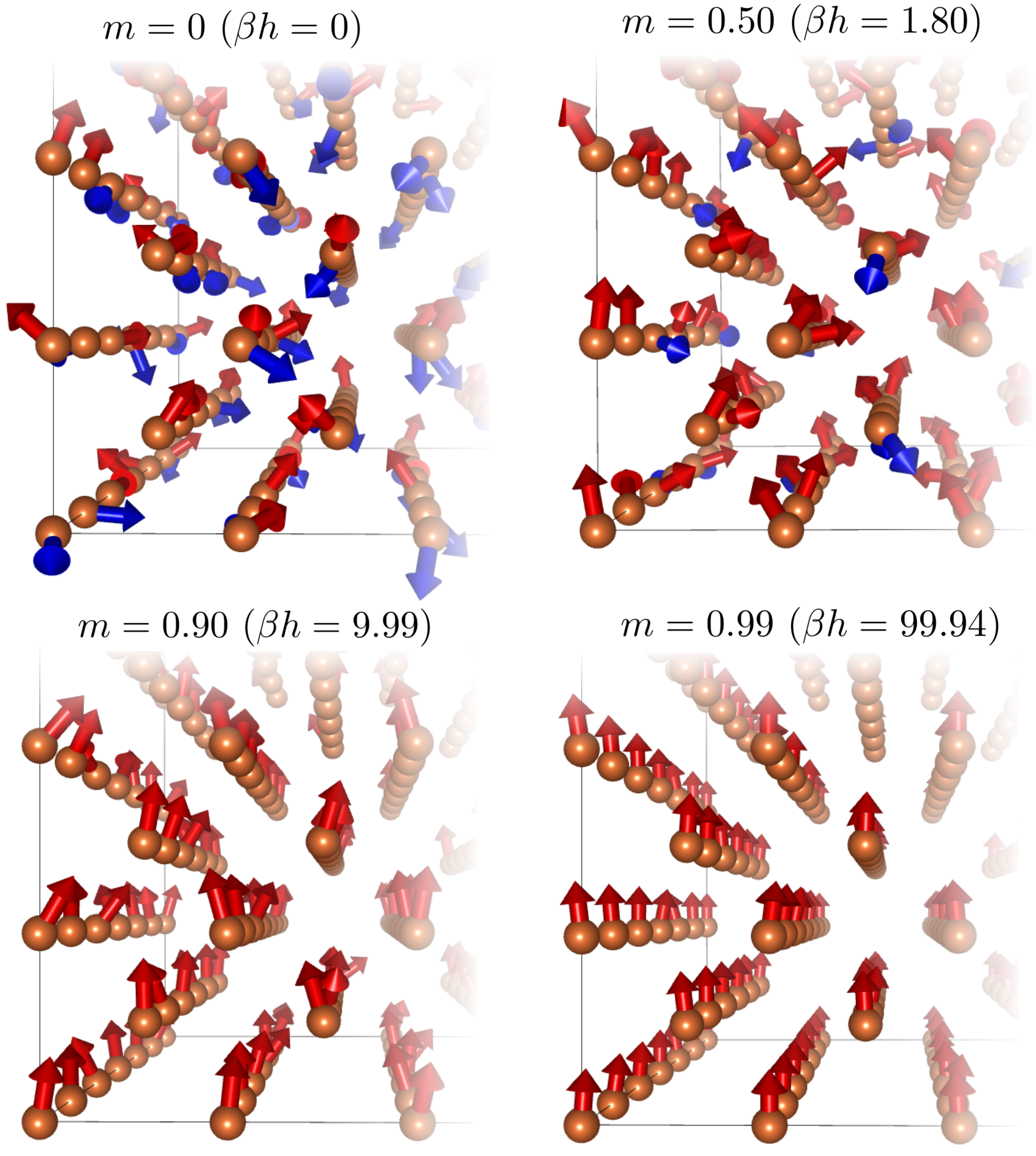}
\caption{
Magnetically constrained supercells for different 
local moment configurations $\{\hat{\textbf{e}}_n\}$, represented by the arrows, for the case study of ferromagnetic bcc Fe. Different temperatures correspond to different orientational probability distributions $P(\{\hat{\textbf{e}}_n\})$.  The method is illustrated here for $P(\{\hat{\textbf{e}}_n\})$  fully prescribed by single-site internal magnetic fields $\{\beta\textbf{h}_n = \beta\textbf{h}\}$, or equivalently by single-site local order parameters $\{\textbf{m}_n = \textbf{m}\}$,
where $\beta\textbf{h}$ is set along $\hat{\textbf{z}}$ (see section \ref{MFA}).
Note that these quantities are site-independent for a ferromagnetic state.
The smaller the value of $m$, the larger the magnetic thermal fluctuations. 
Results are shown for $\{m=0, 0.50, 0.90, 0.99\}$, which correspond to $\{\beta h=0, 1.80, 9.99, 99.94\}$ via Eq.\ (\ref{EQ_mn}).
Red and blue arrows are used for orientations with a polar angle being $\theta_n<\pi/2$ and $\theta_n\geq \pi/2$, respectively.
}%
\label{Fig1a}
\end{figure}

The novel aspect of our work is that the thermodynamic averages to compute $\mathcal{G}_\text{tot}$ are performed as superpositions of supercell \textit{ab initio} calculations that are magnetically constrained to different sets of $\{\hat{\textbf{e}}_n\}$.
Figure \ref{Fig1a} shows magnetically constrained supercells for a thermodynamically excited ferromagnetic state of bcc Fe using this method, where higher and lower temperatures correspond to a set of largely disordered and ordered local moment orientations $\{\hat{\textbf{e}}_n\}$, respectively. 
Most importantly, Eq.\ (\ref{Eq_GIntro}) is a general expression that we can compute for an arbitrary probability distribution. In our approach, therefore, different theories for $P(\{\hat{\textbf{e}}_n\})$ can be constructed by restricting the thermal configurations to a meaningful magnetic phase space described by a trial magnetic Hamiltonian that can include the effect of nonlocal correlations, which will be explained in section \ref{Theory} [see Eq.\ (\ref{Eq_HIntro})].
Another important outcome of this work is the demonstration that the statistics of noncollinear magnetic configurations can be described even with a low number of constrained calculations for supercells of relatively small size, thus being computationally affordable.

A key aspect of our approach is to provide the dependence of $\mathcal{G}_\text{tot}$ on an expansion in powers of magnetic order parameters akin to a Ginzburg-Landau theory. This describes a hierarchy of local moment correlation functions~\cite{PhysRevB.99.144424}, accounting for pairwise and multisite magnetic interactions, and their magnetostructural coupling (see section \ref{MultiMVE}). Higher than pairwise free energy terms can generate discontinuous and ordered-to-ordered magnetic phase transitions, fully quantified and provided by our approach. As a demonstrator of this, we study the noncollinear, geometrically frustrated, magnetism of Mn-based antiperovskite materials~\cite{LB1981}, which are driving strong interest owing to a discontinuous phase transition to a triangular antiferromagnetic state with a giant barocaloric effect~\cite{Matsunami1,PhysRevX.8.041035,TAO2021114049}. We provide the \textit{ab initio} origin of this transition, how it depends on the chemical composition, and compare our results with experiment.

This paper is organized as follows. In section \ref{Theory} we set the theoretical framework and present our approach. Section \ref{Ecalc} explains how to compute Eq.\ (\ref{Eq_UmagIntro}) using magnetically constrained supercells.
In sections \ref{MultiMVE} and \ref{FOT} we describe the effect of multisite magnetic interactions and magnetovolume coupling on the character of magnetic phase transitions. Computational details are given in section \ref{Comp}. In section  \ref{bccFe} we apply our approach to the well studied ferromagnetic state of bcc Fe as an instructive example, while section \ref{Mn3AN} focuses on the magnetism of antiperovskite materials.  Conclusions and outlook are given in section \ref{Conc}. The paper finalizes with appendix \ref{App1}, which shows the performance of the Gibbs free energy calculation using smaller supercells.

%%%%%%%%%%%%%%%%%%%%%%%FIGURE%%%%%%%%%%%%%%%%%%%%%%%%%%%%%%%%%%%%%
%%%%%%%%%%%%%%%%%%%%%%%%%%%%%%%%%%%%%%%%%%%%%%%%%%%%%%%%%%%%%%%%%%%%

%%%%%%%%%%%%%%%%%%%%%%%%%%%%%%%%%%%%%%%%%%%%%%%%%%%%%%%%%%%
%%%%%%%%%%%%%%%%%%%%%%%%%%%%%%%%%%%%%%%%%%%%%%%%%%%%%%%%%%%
\section{\textit{Ab initio} Theory of the magnetic Gibbs free energy}
\label{Theory}
%%%%%%%%%%%%%%%%%%%%%%%%%%%%%%%%%%%%%%%%%%%%%%%%%%%%%%%%%%%
%%%%%%%%%%%%%%%%%%%%%%%%%%%%%%%%%%%%%%%%%%%%%%%%%%%%%%%%%%%

%%%%%%%%%%%%%%%%%%%%%%%%%%%%%%%%%%%%%%%%%%%%%%%%%%%%%%%%%%%
\subsection{Theoretical framework}
\label{Framework}
%%%%%%%%%%%%%%%%%%%%%%%%%%%%%%%%%%%%%%%%%%%%%%%%%%%%%%%%%%%

The basis of our \textit{ab initio} modelling is density functional theory (DFT), formally extendable to finite temperatures~\cite{PhysRev.137.A1441}. DFT describes the local moment orientations $\{\hat{\textbf{e}}_n\}$ as emerging quantities from the local spin-polarization of the magnetic moment density. This establishes a framework to model transverse magnetic fluctuations at different temperatures, at and below the paramagnetic state, known as the disordered local moment (DLM) theory~\cite{0305-4608-15-6-018}.
Our approach is based on this DLM theory and allows its application generally with any \textit{ab initio} machinery able to employ magnetic constrains, as for example developed in Refs.\ \cite{PhysRevB.91.054420,PhysRevB.69.024415,doi:10.1063/1.370494,PhysRevB.102.144101}. Hence, we open the opportunity to use DFT codes based on a plane wave basis instead of the Korringa-Kohn-Rostoker (KKR) formulation of DFT, and its bond to the single-site coherent potential approximation~\cite{PhysRev.156.809,PhysRevB.5.2382} to carry out the ensemble averages.

The central component of DLM theory is an adiabatic approximation for the orientations of the local magnetic moments, $\{\hat{\textbf{e}}_n\}$. These orientations are thus considered to evolve very slowly in comparison with the fast electronic motions forming the underlying many-electron interacting subsystem.
Such an assumption means that thermally-induced transverse magnetic excitations can be obtained from DFT calculations constrained to comply with different magnetic configurations $\{\hat{\textbf{e}}_n\}$, i.e.,
\begin{equation}
    \int_{\Omega_n} d^3 \textbf{r} \boldsymbol{\mu}(\textbf{r};\{\hat{\textbf{e}}_n\})
    =\mu_n(\{\hat{\textbf{e}}_n\})\hat{\textbf{e}}_n,
    \label{EQ_Muconstr}
\end{equation}
where $\boldsymbol{\mu}(\textbf{r};\{\hat{\textbf{e}}_n\})$ is the magnetic moment density, and $\Omega_n$ is the volume defining the region at site $n$ in which a local moment of size $\mu_n$ emerges.

Formally, a self-consistent calculation of the DFT-based total energy for given magnetic configurations $\{\hat{\textbf{e}}_n\}$ can be used to obtain the magnetic energy, $E_\text{int}(\{\hat{\textbf{e}}_n\})$~\cite{0305-4608-15-6-018}, introduced in Eq.\ (\ref{Eq_UmagIntro}). Such a calculation should be available by, for example, applying constraining single-site magnetic fields that establish $\{\hat{\textbf{e}}_n\}$ as minima of the system.
We consider that the magnitudes $\{\mu_n\}$ evolve in a much shorter timescale than the local moment orientations so that $\{\mu_n\}$ instantaneously adapt to a given configuration $\{\hat{\textbf{e}}_n\}$. We thus obtain $\{\mu_n\}$ self-consistently with respect to a given orientational configuration $\{\hat{\textbf{e}}_n\}$. However, their thermal fluctuations could be described by performing additional averages~\cite{PhysRevB.95.054426}.

$E_\text{int}(\{\hat{\textbf{e}}_n\})$ provides the corresponding partition function
\begin{equation}
    \mathcal{Z}=\int \prod_n\left[d\hat{\textbf{e}}_n\right]
    \exp\left[-\beta E(\{\hat{\textbf{e}}_n\})\right],
    \label{EQ_Z}
\end{equation}
where $\beta=\frac{1}{k_\text{B} T}$, $k_\text{B}$ and $T$ being the Boltzmann constant and the temperature, and
\begin{equation}
    E(\{\hat{\textbf{e}}_n\})
    = E_\text{int}(\{\hat{\textbf{e}}_n\})-\textbf{H}\cdot\sum_n \mu_n\hat{\textbf{e}}_n
    \label{EQ_Eint}
\end{equation}
is the magnetic energy also including a coupling of the local magnetic moments with an external magnetic field $\textbf{H}$.
Note that $\{\hat{\textbf{e}}_n\}$ are treated as classical quantities and so $\mathcal{Z}$ and ensemble averages are obtained by performing integrals over all solid angle space, $d\hat{\textbf{e}}_n=d\theta_n d\phi_n\sin\theta_n$. The probability of the system being in a magnetic state with local moment orientations $\{\hat{\textbf{e}}_n\}$ is, therefore,
\begin{equation}
    P(\{\hat{\textbf{e}}_n\})=\frac{\exp\left[-\beta E(\{\hat{\textbf{e}}_n\})\right]}{\mathcal{Z}},
    \label{EQ_P}
\end{equation}
which gives the exact magnetic Gibbs free energy of the system by carrying out ensemble averages over all appropriately weighted magnetic configurations,
\begin{equation}
\begin{split}
   \mathcal{G} &
    =-k_\text{B} T\ln\mathcal{Z} \\
   & =\int \prod_n\left[d\hat{\textbf{e}}_n\right] P(\{\hat{\textbf{e}}_n\})
    \Big[
    E(\{\hat{\textbf{e}}_n\})
    +k_\text{B} T\ln P(\{\hat{\textbf{e}}_n\})
    \Big] .
    \label{EQ_G}
\end{split}
\end{equation}

The complex many-electron origins of $E_\text{int}(\{\hat{\textbf{e}}_n\})$ in metallic systems makes its dependence on the orientational magnetic configurations very complicated. Consequently, $E_\text{int}(\{\hat{\textbf{e}}_n\})$ can contain pairwise as well as higher order magnetic interaction terms,
\begin{equation}
\begin{split}
    E_\text{int}(\{\hat{\textbf{e}}_n\})
    = & E_\text{int,0}
    -\sum_{ij}J_{ij} \hat{\textbf{e}}_{i}\cdot\hat{\textbf{e}}_{j} \\
   & - \sum_{ijkl}K_{ijkl} (\hat{\textbf{e}}_{i}\cdot\hat{\textbf{e}}_{j})
    (\hat{\textbf{e}}_{k}\cdot\hat{\textbf{e}}_{l})
    -\dots,
    \label{EQ_Egen}
\end{split}
\end{equation}
where $E_\text{int,0}$ is a reference energy. In principle, the full complexity and myriad of magnetic interactions in $E_\text{int}(\{\hat{\textbf{e}}_n\})$ could be computed using DFT calculations if given enough computer time.
However, the necessary computational costs are very far from available high-performance computational resources.
Alternatively, we use a trial Hamiltonian generally given as
%%%%%%%%%%%%%%%%%%%%%%%%%%%%%%%%%%%%%%%%%%%%%%%%%%%%%%%%%%%
\begin{equation}
\begin{split}
\mathcal{H}_\text{tr}= & 
-\sum_n\textbf{h}_n\cdot\hat{\textbf{e}}_n
%+H^{(2)}(\hat{\textbf{e}}_n,\hat{\textbf{e}}_{n'}) \\
+\sum_{n,n'}H^{(2)}_{n,n'}(\hat{\textbf{e}}_n,\hat{\textbf{e}}_{n'}) \\
& +\sum_{n,n',n''}H^{(3)}_{n,n',n''}(\hat{\textbf{e}}_n,\hat{\textbf{e}}_{n'},\hat{\textbf{e}}_{n''})
%& +H^{(3)}(\hat{\textbf{e}}_n,\hat{\textbf{e}}_{n'},\hat{\textbf{e}}_{n''})
+\cdots
,
\end{split}
\label{Eq_HIntro}
\end{equation}
%%%%%%%%%%%%%%%%%%%%%%%%%%%%%%%%%%%%%%%%%%%%%%%%%%%%%%%%%%%
where $\{\textbf{h}_n\}$ are single-site internal magnetic fields, also called Weiss fields, forming the simplest mean-field theory for the interactions.
In this work we only consider the single-site terms [see Eq.\ (\ref{EQ_H0})], but
by including higher order terms, $\{H^{(2)}_{n,n'}(\hat{\textbf{e}}_n,\hat{\textbf{e}}_{n'}),\dots\}$, our treatment can be systematically improved to account for different types of nonlocal magnetic correlations~\cite{0305-4608-15-6-018}.
In section \ref{MFA} we invoke the Peierls-Feynman inequality to find an upper-bound of $\mathcal{G}$~\cite{PhysRev.97.660,doi:10.1143/JPSJ.50.1854}, in analogy to the Gibbs-Bogoliubov inequality typically applied for the free energy of an ionic Hamiltonian~\cite{Isihara_1968,PhysRevB.89.064302}. This establishes a theory to calculate the Gibbs free energy and $U_\text{mag}=\langle E_\text{int}\rangle_\text{tr}$, which contains the effect of all magnetic interactions, with respect to the associated trial probability distribution,
\begin{equation}
    P(\{\hat{\textbf{e}}_n\})\approx
    P_\text{tr}(\{\hat{\textbf{e}}_n\})=\frac{1}{\mathcal{Z}_\text{tr}}\exp\left[-\beta\mathcal{H}_\text{tr}\right],
    \label{EQ_P0gen}
\end{equation}
where $\mathcal{Z}_\text{tr}$ is the corresponding partition function.
%We also highlight that from Eq.\ (\ref{EQ_Egen}) one can see that the presence of quartic and higher order magnetic interactions implies that different values of effective pairwise interactions are obtained when these are computed in different referent magnetic states.

{\renewcommand{\arraystretch}{1.5}
\begin{table*}[t]
    \centering
    \begin{tabular}{ll}
    \hline\hline
        1. & A set of values  $\{\textbf{m}_n\}=\{\textbf{m}_1,\dots,\textbf{m}_N\}$ are chosen, $N$ being the number of magnetically constrained sites in the supercell. \\
        2. & From $\{\textbf{m}_n\}$, we obtain $P_\text{tr}(\{\hat{\textbf{e}}_n\})$ using equations (\ref{EQ_mn}) and (\ref{EQ_Pn_2}) via mapping through $\{\beta\textbf{h}_n\}$. See Sec.\ \ref{MFA} for detail. \\
        3. & $N_\text{MC}$ DFT calculations with local moment orientations $\{\hat{\textbf{e}}_n\}$ constrained following $P_\text{tr}(\{\hat{\textbf{e}}_n\})$ are performed. Sect.\ \ref{configurations} \\
          &   gives detail on how to obtain $\{\hat{\textbf{e}}_n\}$ for each DFT calculation, which correspond to different polar and azimuthal angles  \\
          & given by equations (\ref{EQ_theta}) and (\ref{EQ_phi}), $(\{\theta_n,\phi_n\}_1,\{\theta_n,\phi_n\}_2,\dots,\{\theta_n,\phi_n\}_{N_\text{MC}}$).  \\
        4. & The internal magnetic energy is computed by carrying out the average $U_\text{mag}=\langle E_\text{int}\rangle_\text{tr}$, using Eq.\ (\ref{EQ_AvEint_mc}) in section \ref{Ecalc}. \\
        5. & Steps 1-4 are repeated for different sets of values of $\{\textbf{m}_n\}$ to obtain $U_\text{mag}(\{\textbf{m}_n\})$. This is done for $\{\textbf{m}_n\}$ describing the  \\
          & magnetic phases of interest. For example, in the case of the ferromagnetic state of bcc Fe the magnitudes of the order \\
          & parameters are identical at all sites. Different values spanning from $m_n=0$ (paramagnetic) to $m_n=1$ (fully ordered \\
          & ferromagnetic) can then be chosen for a given spin-polarization axis.
        \\
    \hline\hline
    \end{tabular}
    \caption{Central steps to obtain the dependence of the internal magnetic energy on the local order parameters.}
    \label{tab:scheme}
\end{table*}
}

The consideration of Eq.\ (\ref{EQ_P0gen}) reduces the thermally induced magnetic configurations to be studied to an affordable and meaningful magnetic phase space. A central task of our approach is to provide the internal magnetic energy as a function of the local magnetic order parameters of the system for this probability distribution,
\begin{equation}
    \textbf{m}_n = \langle \hat{\textbf{e}}_n\rangle_\text{tr},
    \label{EQ_mn_0}
\end{equation}
i.e.\ $U_\text{mag}(\{\textbf{m}_n\})$.
$\langle\cdots\rangle_\text{tr}$ indicates an average over $\{\hat{\textbf{e}}_n\}$ with respect to $P_\text{tr}(\{\hat{\textbf{e}}_n\})$.
As will be explained in section \ref{MFA} and as illustrated in figure \ref{Fig1a}, in this work the trial orientational probability distribution is prescribed by the values of the order parameters, $P_\text{tr}(\{\hat{\textbf{e}}_n\};\{\textbf{m}_n\})$. In Table \ref{tab:scheme} we advance and summarize the central steps, and their relation to the different sections and equations in the paper, for the computation of $U_\text{mag}(\{\textbf{m}_n\})$.
These steps can be repeated for different lattice attributes to also obtain the dependence on the crystal structure and the consequent magnetostructural coupling.
Once $U_\text{mag}$ and its dependencies have been computed, the Gibbs free energy is directly obtainable from Eq.\ (\ref{Eq_GIntro}).

%%%%%%%%%%%%%%%%%%%%%%%%%%%%%%%%%%%%%%%%%%%%%%%%%%%%%%%%%%%
\subsection{Computation of $U_\text{mag}=\langle E_\text{int}\rangle_\text{tr}$ using magnetically constrained supercell calculations}
\label{Ecalc}
%%%%%%%%%%%%%%%%%%%%%%%%%%%%%%%%%%%%%%%%%%%%%%%%%%%%%%%%%%%

$U_\text{mag}=\langle E_\text{int}\rangle_\text{tr}$ is obtained by carrying out an average of the magnetic energy over $\{\hat{\textbf{e}}_n\}$ with respect to a trial probability distribution $P_\text{tr}(\{\hat{\textbf{e}}_n\})$ through Eq.\ (\ref{Eq_UmagIntro}).
Formally, a Monte Carlo integration can be employed,
\begin{equation}
    \langle E_\text{int}\rangle_\text{tr}\approx
     \frac{V_{\Omega}^{N_\text{MC}}}{N_\text{MC}}
    \sum_i^{N_\text{MC}}
    \left[
    %\prod_n\left[ P_n(\hat{\textbf{e}}_n)\right]_{\{\hat{\textbf{e}}_n\}_i^\text{rand}}
    P_\text{tr}(\{\hat{\textbf{e}}_n\}_i^\text{rand})
    E_\text{int}(\{\hat{\textbf{e}}_n\}_i^\text{rand})
    \right],
    \label{EQ_AvEint_MC}
\end{equation}
where $V_{\Omega}=\int d\hat{\textbf{e}}_n=4\pi$ is the solid angle volume, $\{\hat{\textbf{e}}_n\}_i^\text{rand}$ is a set of fully random local moment orientations, and $N_\text{MC}$ is the number of sets generated. $E_\text{int}(\{\hat{\textbf{e}}_n\}_i^\text{rand})$ can be calculated using constraining magnetic fields in DFT calculations for supercells  with fully random local moments. The number of atoms within the supercell, $N_\text{sc}$, must be large enough to reduce box-size effects satisfactorily. Once $N_\text{MC}$ magnetically constrained supercell calculations have been performed, Eq.\ (\ref{EQ_AvEint_MC}) provides an approximated value of $\langle E_\text{int}\rangle_\text{tr}$ for an arbitrary shape of $P_\text{tr}(\{\hat{\textbf{e}}_n\})$.
Note that different $P_\text{tr}(\{\hat{\textbf{e}}_n\})$ correspond to different states of magnetic thermal disorder, see Fig.\ \ref{Fig1a}. 
However, the minimum value of $N_\text{MC}$ to obtain accurately enough results increases for probability distributions $P_\text{tr}(\{\hat{\textbf{e}}_n\})$ that describe magnetic configurations away from fully random local moment orientations. It is to be presumed, therefore, that $N_\text{MC}$ is prohibitively large.

To circumvent this computational constrain, we approximate Eq.\ (\ref{Eq_UmagIntro}) instead by
\begin{equation}
    \langle E_\text{int}\rangle_\text{tr}\approx
    \frac{1}{N_\text{MC}}\sum_i^{N_\text{MC}}
    E_\text{int}\left(\{\hat{\textbf{e}}_n\}_i^{P_\text{tr}(\{\hat{\textbf{e}}_n\})}\right),
    \label{EQ_AvEint_mc}
\end{equation}
where now $N_\text{MC}$ magnetically constrained DFT calculations are performed for sets of local moment orientations that, instead of being fully random, directly obey the probability distribution
We have found (see sections \ref{bccFe} and \ref{Mn3AN}) that computationally affordable values of $N_\text{MC}$ can be achieved using Eq.\ (\ref{EQ_AvEint_mc}) because the constrained magnetic configurations here already follow $P_\text{tr}(\{\hat{\textbf{e}}_n\})$.
The downside, however, is that a whole set of $N_\text{MC}$ configurations needs to be computed for each probability distribution that one aims to study, for example for each panel in Fig.\ \ref{Fig1a}.
In the following section we show  the case of the simplest single-site mean-field theory.

%%%%%%%%%%%%%%%%%%%%%%%%%%%%%%%%%%%%%%%%%%%%%%%%%%%%%%%%%%%
\subsubsection{Peierls-Feynman inequality}
\label{MFA}
%%%%%%%%%%%%%%%%%%%%%%%%%%%%%%%%%%%%%%%%%%%%%%%%%%%%%%%%%%%

Albeit our supercell approach to compute Eq.\ (\ref{EQ_AvEint_mc}) is in principle implementable for a general expression of Eq.\ (\ref{Eq_HIntro}), we choose a trial Hamiltonian that contains only single-site magnetic fields, or Weiss fields, as originally used by Gy\H{o}rffy \textit{et al}.~\cite{0305-4608-15-6-018},
\begin{equation}
    \mathcal{H}_\text{tr}=-\sum_n\textbf{h}_n\cdot\hat{\textbf{e}}_n.
    \label{EQ_H0}
\end{equation}
This $\mathcal{H}_\text{tr}$ describes local magnetic moments whose orientations are sustained by mean fields $\{\textbf{h}_n\}$, which can contain the effect of an external magnetic field if present. The partition function associated with $\mathcal{H}_\text{tr}$ is
\begin{equation}
\begin{split}
    \mathcal{Z}_\text{tr} &
    =\prod_n\int d\hat{\textbf{e}}_n \exp
    \left[
    \beta\textbf{h}_n\cdot\hat{\textbf{e}}_n
    \right]
    =\prod_n 4\pi \frac{\sinh(\beta h_n)}{\beta h_n},
    \label{EQ_Z0}
\end{split}
\end{equation}
and so the corresponding probability for a magnetic configuration $\{\hat{\textbf{e}}_n\}$ becomes
\begin{equation}
    P_\text{tr}(\{\hat{\textbf{e}}_n\})=\frac{1}{\mathcal{Z}_\text{tr}}\prod_n\exp\left[\beta\textbf{h}_n\cdot\hat{\textbf{e}}_n\right]=\prod_n P_n(\hat{\textbf{e}}_n),
    \label{EQ_P0}
\end{equation}
where $h_n$ is the magnitude of $\textbf{h}_n$. Owing to the single-site nature of $\mathcal{H}_\text{tr}$, Eq.\ (\ref{EQ_P0}) defines single-site probability distributions for each local moment orientation,
\begin{equation}
    P_n(\hat{\textbf{e}}_n)=\frac{\beta h_n}{4\pi\sinh(\beta h_n)} \exp\left[\beta\textbf{h}_n\cdot\hat{\textbf{e}}_n\right].
    \label{EQ_Pn}
\end{equation}

The Peierls-Feynman inequality~\cite{PhysRev.97.660,doi:10.1143/JPSJ.50.1854} provides an upper bound of $\mathcal{G}$ given in Eq.\ (\ref{EQ_G}), which we refer to as $\mathcal{G}_u$,
\begin{equation}
    \mathcal{G}\leq
    \mathcal{G}_u = \mathcal{G}_\text{tr}
    +\langle E(\{\hat{\textbf{e}}_n\})\rangle_\text{tr}
    -\langle\mathcal{H}_\text{tr}(\{\hat{\textbf{e}}_n\})
    \rangle_\text{tr},
    \label{EQ_PF}
\end{equation}
where $\mathcal{G}_\text{tr}=-k_\text{B}T \ln\mathcal{Z}_\text{tr}$ is the associated Gibbs free energy of the trial Hamiltonian, and we recall that $\langle\cdots\rangle_\text{tr}$ indicates an average over $\{\hat{\textbf{e}}_n\}$ with respect to $P_\text{tr}(\{\hat{\textbf{e}}_n\})$. 
From Eq.\ (\ref{EQ_Eint}) and Eq.\ (\ref{EQ_PF}) one can finally write
\begin{equation}
    \mathcal{G}_u
    =\langle E_\text{int}\rangle_\text{tr}
    -\textbf{H}\cdot\sum_n \mu_n\textbf{m}_n
    -TS_\text{mag}.
    \label{EQ_Gu}
\end{equation}
The previous result becomes Eq.\ (\ref{Eq_GIntro}) after applying a Legendre transform accounting for the effect of mechanical stress.
It also introduces the magnetic local order parameters as the averages of the local moment orientations, as previously given in Eq.\ (\ref{EQ_mn_0}),
\begin{equation}
\begin{split}
    \textbf{m}_n=\langle\hat{\textbf{e}}_n\rangle_\text{tr}
    & =\int d\hat{\textbf{e}}_n P_n(\hat{\textbf{e}}_n)\hat{\textbf{e}}_n\\
    & =\left[
    \frac{-1}{\beta h_n}+\coth(\beta h_n)
    \right]\hat{\textbf{h}}_n,
\end{split}
    \label{EQ_mn}
\end{equation}
as well as an expression for the magnetic entropy associated with the orientational configurations $\{\hat{\textbf{e}}_n\}$,
\begin{equation}
    S_\text{mag}(\{\textbf{m}_n\})=\sum_n S_n(\textbf{m}_n)
    \label{EQ_Smag}
\end{equation}
where
\begin{equation}
\begin{split}
    S_n
  &  =-k_\text{B}\int d\hat{\textbf{e}}_n P_n(\hat{\textbf{e}}_n)\ln P_n(\hat{\textbf{e}}_n) \\
  &  =k_\text{B}
    \left[
    1+\ln\left(4\pi\frac{\sinh(\beta h_n)}{\beta h_n}\right)
    -\beta h_n\coth(\beta h_n)
    \right].
\end{split}
    \label{EQ_Sn}
\end{equation}
Note that $\{P_n(\hat{\textbf{e}}_n)\}$ is prescribed by $\{\textbf{m}_n\}$ since both of them are univocally given by $\{\beta\textbf{h}_n\}$.

\begin{figure}[t]
\centering
\includegraphics[clip,scale=0.22]{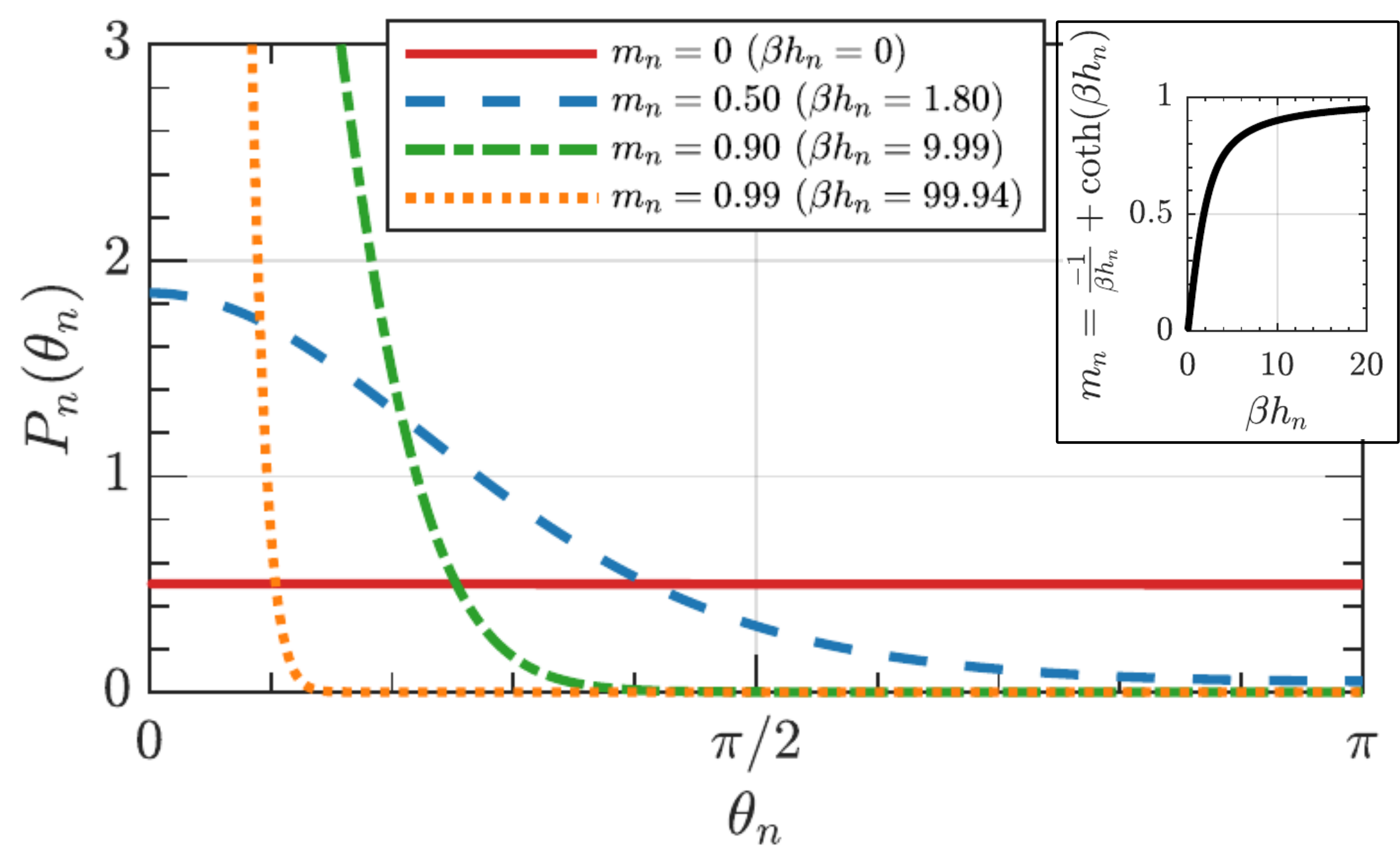}
\caption{
Single-site probability distribution
for different values of $m_n$ [see Eq.\ (\ref{EQ_Pn_2})].
The inset shows how the dependence of $m_n$ on $\beta h_n$ is described by the Langevin function [see Eq.\ (\ref{EQ_mn})].
}%
\label{Fig1b}
\end{figure}

This mean-field theory produces a typical classical behavior of $\textbf{m}_n$ following the Langevin function, as shown in Eq.\ (\ref{EQ_mn}) and plotted in the inset of Fig.\ \ref{Fig1b}. At zero temperature the local order parameter, which is always parallel to $\hat{\textbf{h}}_n$, has a magnitude equal to one, $\textbf{m}_n=\hat{\textbf{h}}_n$. 
This means that $\hat{\textbf{e}}_n$ does not thermally fluctuate.
Note that in this limit, $T\rightarrow 0$, $\{\beta h_n\}\rightarrow \infty$ and $\{h_n\}$ has a finite value.
On the other hand, at high enough temperatures the thermal fluctuations can become sufficiently large to establish local moments fluctuating randomly over all possible orientations. This corresponds to a fully disordered, paramagnetic, state described by $\textbf{m}_n=\textbf{0}$ ($\{h_n=0\}$) and being present above Curie or N\'{e}el transition temperatures. 
We highlight that these transition temperatures as well as the dependence of the order parameters on temperature, $\{\textbf{m}_n(T)\}$, are obtained by minimizing the free energy at different values of $T$, as explained in section \ref{FOT}.
%See section \ref{Ecalc} and Fig.\ \ref{Fig1} for further detail.
The snapshots shown in Fig.\ \ref{Fig1a} of different magnetically constrained configurations correspond to different probabilities $\{P_n(\hat{\textbf{e}}_n)\}$ (or $\{\textbf{m}_n\}$) for a thermally fluctuating ferromagnetic state of bcc Fe.

%%%%%%%%%%%%%%%%%%%%%%%FIGURE%%%%%%%%%%%%%%%%%%%%%%%%%%%%%%%%%%%%%
%%%%%%%%%%%%%%%%%%%%%%%%%%%%%%%%%%%%%%%%%%%%%%%%%%%%%%%%%%%%%%%%%%%%

%%%%%%%%%%%%%%%%%%%%%%%%%%%%%%%%%%%%%%%%%%%%%%%%%%%%%%%%%%%
\subsubsection{Magnetically constrained supercell configurations}
\label{configurations}
%%%%%%%%%%%%%%%%%%%%%%%%%%%%%%%%%%%%%%%%%%%%%%%%%%%%%%%%%%%

As indicated above, the probability distribution at each magnetic site $n$ follows Eq.\ (\ref{EQ_Pn}) independently owing to the single-site nature of Eq.\ (\ref{EQ_H0}),
\begin{equation}
    P_n(\hat{\textbf{e}}_n)=\frac{\beta h_n}{4\pi\sinh(\beta h_n)} \exp\left[\beta h_n \cos\theta_n\right],
    \label{EQ_Pn_2}
\end{equation}
where $0\leq\theta_n\leq\pi$ is the angle between $\hat{\textbf{e}}_n$ and $\textbf{h}_n$ (or the magnetic local order parameter $\textbf{m}_n$).
In Fig.\ \ref{Fig1b} we show the behavior of $P_n(\hat{\textbf{e}}_n)$ against $\theta_n$ for different values of $\beta h_n$. While $\beta h_n=0$ describes a magnetic site that is fully disordered ($m_n=0$), $\beta h_n\rightarrow\infty$ corresponds to a non-fluctuating local moment orientation $\hat{\textbf{e}}_n=\hat{\textbf{h}}_n$ ($m_n=1$).
We use a standard generator of uniformly distributed random numbers to obtain different magnetic configurations, which are the inputs in our magnetically constrained supercell calculations.
Rotating the $\hat{\textbf{z}}$-axis such that it is aligned with $\textbf{h}_n$ allows one to write $\hat{\textbf{e}}_n=(\sin\theta_n\cos\varphi_n,\sin\theta_n\sin\varphi_n,\cos\theta_n)$, where $0\leq\varphi_n\leq 2\pi$ is an in-plane angle.
Since the probability integral must be normalized, we can obtain $\theta_n$ by considering
\begin{equation}
    2\pi\int_0^{\theta_n} d\theta \sin\theta \frac{\beta h_n}{4\pi\sinh(\beta h_n)} \exp\left[\beta h_n \cos\theta\right]
    =x_n^{\theta},
    \label{EQ_thetaInt}
\end{equation}
where $x_n^{\theta}$ is a random number ($\{0 \leq x_n^{\theta}\leq 1\}$) and the $2\pi$ factor comes from the in-plane integration. Solving Eq.\ (\ref{EQ_thetaInt}) gives
\begin{equation}
    {\theta_n}=\arccos\left\{
    \frac{1}{\beta h_n}\left(
    \ln\Big[\exp(\beta h_n)-2x_n^{\theta}\sinh(\beta h_n)\Big]
    \right)
    \right\}.
    \label{EQ_theta}
\end{equation}
On the other hand, Eq.\ (\ref{EQ_Pn_2}) does not depend on $\varphi_n$, which means that $\varphi_n$ is fully random and can be obtained directly using another random number $x_n^{\varphi}$ ($\{0 \leq x_n^{\varphi}\leq 1\}$),
\begin{equation}
    \varphi_n=2\pi x_n^{\varphi}.
    \label{EQ_phi}
\end{equation}
The local moment orientations for each magnetic configuration $\{\hat{\textbf{e}}_n\}_i^{\{P_n(\hat{\textbf{e}}_n)\}}$ are thus obtained from $N$ pairs of random numbers, $\{x_n^{\theta},x_n^{\varphi}\}$, using Eq.\ (\ref{EQ_theta}) and Eq.\ (\ref{EQ_phi}) for given values of $\{\beta\textbf{h}_n\}$ (or $\{\textbf{m}_n\}$) defining the probability distribution. Here $N$ is the number of magnetic sites in the supercell. This is how the orientations shown in Fig.\ \ref{Fig1a} have been obtained.

%%%%%%%%%%%%%%%%%%%%%%%%%%%%%%%%%%%%%%%%%%%%%%%%%%%%%%%%%%%
\subsection{Hierarchy of local moment correlation functions and their magnetoelastic coupling}
\label{MultiMVE}
%%%%%%%%%%%%%%%%%%%%%%%%%%%%%%%%%%%%%%%%%%%%%%%%%%%%%%%%%%%

Since in our mean-field theory $P_\text{tr}(\{\hat{\textbf{e}}_n\})$ is uniquely prescribed by $\{\beta\textbf{h}_n\}$, which in turn unequivocally map to $\{\textbf{m}_n\}$, the average of $E_\text{int}$ has a direct dependence on $\{\textbf{m}_n\}$. From Eq.\ (\ref{Eq_UmagIntro}) and Eq.\ (\ref{EQ_Egen}) we, therefore, have
\begin{equation}
\begin{split}
    U_\text{mag}= & \langle E_\text{int}\rangle_\text{tr}=
    U^{(0)}
  -\sum_{ij}U^{(2)}_{ij}\textbf{m}_{i}\cdot\textbf{m}_{j} \\
   & -\sum_{ijkl}U^{(4)}_{ijkl}(\textbf{m}_{i}\cdot\textbf{m}_{j})(\textbf{m}_{k}\cdot\textbf{m}_{l})
    -\text{h.o.},
    \label{EQ_Eintmn}
\end{split}
\end{equation}
where $\{U^{(2)}_{ij}, U^{(4)}_{ijkl}, \dots\}$ form a hierarchy of second and higher order local moment correlations that are characteristic of the magnetic material investigated. Here $\text{h.o.}$ stands for higher order terms. These local moment correlations describe expansion coefficients of the internal magnetic energy in terms of the magnetic local order parameters, in the spirit of a Ginzburg-Landau theory. As mentioned above, the dependence of $\langle E_\text{int}\rangle$ on $\{\textbf{m}_n\}$ is performed through the computation of Eq.\ (\ref{EQ_AvEint_mc}) using magnetically constrained supercells. We realize this calculation for a reduced phase space of $\{\textbf{m}_n\}$ containing the magnetic states of interest only, e.g.\ the ferromagnetic state in bcc Fe and a triangular antiferromagnetic state in section \ref{Mn3AN}. See Table \ref{tab:scheme}.

$\langle E_\text{int}\rangle_\text{tr}$ can be calculated for different lattice parameters and crystal structures and so its dependence on deformation can be also obtained. An expression of the Gibbs free energy in Eq.\ (\ref{EQ_Gu}) that accounts for the effect of an applied hydrostatic pressure $p$ is given by the following Legendre transform
\begin{equation}
\begin{split}
    \mathcal{G}_\text{tot} = & \mathcal{G}_u
    +pV \\
   = & \langle E_\text{int}\rangle_\text{tr}(\{\textbf{m}_n\},V) -TS_\text{mag}(\{\textbf{m}_n\}) \\
    & -\textbf{H}\cdot\sum_n \mu_n\textbf{m}_n
    +pV.
    \label{Eq_Ggenvare}
\end{split}
\end{equation}
Here $\{U^{(0)},U^{(2)}_{ij},U^{(4)}_{ijkl},\cdots\}$ inside $\langle E_\text{int}\rangle_\text{tr}$  depend on the volume $V$.
Considering a parabolic dependence on $V$ for the internal magnetic energy in the paramagnetic limit we can write
\begin{equation}
    \lim_{\{\textbf{m}_n\}\rightarrow\{\textbf{0}\}}\langle E_\text{int}\rangle_\text{tr} = U^{(0)}(\omega)=U^{(0)}(V_\text{PM})+\frac{1}{2}\gamma V_\text{PM}\omega^2,
    \label{Eq_E0vare}
\end{equation}
where $V_\text{PM}$ is the volume that minimizes the free energy in this paramagnetic limit, $\omega=\frac{V-V_\text{PM}}{V_\text{PM}}$ is the relative volume change, and $\gamma$ is the inverse of the compressibility (bulk modulus) describing the energy cost caused by the application of $p$ in the paramagnetic state. $\gamma$ is computed, therefore, by calculating the change of the internal magnetic energy $U_\text{mag}=\langle E_\text{int}\rangle_\text{tr}$ at different volumes in supercells containing fully random local moment orientations. Note that Eq.\ (\ref{Eq_Ggenvare}) can be formally generalized to account for more complicated deformations caused by the application of different types of stresses, as given by Eq.\ (\ref{Eq_GIntro}).

%%%%%%%%%%%%%%%%%%%%%%%%%%%%%%%%%%%%%%%%%%%%%%%%%%%%%%%%%%%
\subsection{Minimization of the Gibbs free energy: First-order magnetic phase transitions generated by electronic and/or magnetoelastic mechanisms}
\label{FOT}
%%%%%%%%%%%%%%%%%%%%%%%%%%%%%%%%%%%%%%%%%%%%%%%%%%%%%%%%%%%

Once $\mathcal{G}_\text{tot}$ has been obtained for a given magnetic material, it can be minimized with respect to $\{\textbf{m}_n\}$ for different values of $T$, $\textbf{H}$, and the crystal structure.
This enables the \textit{ab initio} study of temperature-dependent magnetic anisotropy~\cite{PhysRevB.74.144411,PhysRevLett.120.097202}, magnetic phase transitions and diagrams of different sort~\cite{HughesNat,PhysRevB.89.054427,PhysRevLett.115.207201,PhysRevLett.118.197202,PhysRevB.95.184438}, and (multi-)caloric effects for solid-state refrigeration~\cite{doi:10.1063/5.0003243,PhysRevB.101.174437,PhysRevX.8.041035}. Note that the second term in the right hand side of Eq.\ (\ref{EQ_Gu}), i.e.\ the magnetic entropy, follows analytical expressions that are directly provided by the chosen trial Hamiltonian [see Eq.\ (\ref{EQ_Smag})]. $U_\text{mag}=\langle E_\text{int}\rangle_\text{tr}$ is in consequence the only part of $\mathcal{G}_\text{tot}$ that is material-dependent, obtained from first-principles through the computation of Eq.\ (\ref{EQ_AvEint_mc}).

Higher than quadratic free energy coefficients in Eq.\ (\ref{EQ_Eintmn}) fundamentally generate discontinuous (first-order) magnetic phase transitions from the paramagnetic state. This is a well known result formerly studied in MnAs, for which the emergence of fourth-order coefficients is driven by a magnetovolume coupling~\cite{PhysRev.130.1347} [we also show this here through Eq.\ (\ref{EQ_Gusimp2})].
However, our Gibbs free energy can also depend on quartic and higher order local moment correlation functions ($\{U^{(4)}_{ijkl}(\omega),\dots\}$). These correlations are higher order free energy terms with a purely electronic origin, evidenced by the fact that they can emerge for calculations where the crystal structure is fixed. They describe the feedback to the magnetic interactions from the response of the underlying electronic structure to different states of thermally fluctuating magnetic order $\{\textbf{m}_n\}$~\cite{PhysRevB.99.144424}. We have already shown how these multisite interactions generate first-order magnetic phase transitions experimentally observed in several materials~\cite{PhysRevLett.118.197202,PhysRevX.8.041035,PhysRevB.101.174437}.
An itinerant electron metamagnetism producing first-order transitions was already described by Wolfarth and Rhodes~\cite{doi:10.1080/14786436208213848}, and examples have been reported and explained by Fujita \textit{et al}.\ for La(Fe$_x$Si$_{1-x}$)$_{13}$ compounds~\cite{doi:10.1063/1.370471,OHTA2005431}.

To exemplify how a first-order transition can be produced by both purely electronic and/or magnetoelastic effects, we consider materials that are composed of equivalent magnetic lattice sites. This implies that $\{\textbf{m}_n=m\hat{\textbf{m}}_n\}$, i.e.\ the size of the local order parameters, $m$, is site-independent. In this case and in the absence of external stimuli ($\textbf{H}=0$, $p=0$), the Gibbs free energy in Eq.\ (\ref{Eq_Ggenvare}) becomes
\begin{equation}
\begin{split}
    \mathcal{G}_\text{tot}
   & = -TS_\text{mag}+ U^{(0)}(V_\text{PM})+\frac{1}{2}V_\text{PM}\gamma\omega^2 \\
   & -U^{(2)}(\omega)m^2-U^{(4)}(\omega)m^4
   -U^{(6)}(\omega)m^6-\cdots
    \label{EQ_Gusimp}
\end{split}
\end{equation}
where
\begin{equation}
\begin{split}
   & U^{(2)}(\omega)=\sum_{ij}U^{(2)}_{ij}(\omega)\cos(\theta_{ij}) \\
   & U^{(4)}(\omega)=\sum_{ijkl}U^{(4)}_{ijkl}(\omega)\cos(\theta_{ij})\cos(\theta_{kl}) \\
   & \hspace{2cm}\vdots
    \label{EQ_S2simp}
\end{split}
\end{equation}
play the role of effective coefficients compactly containing the overall effect of all the local moment correlations, $\theta_{ij}$ being the relative angle between $\textbf{m}_i$ and $\textbf{m}_j$.
We now assume that the magnetovolume coupling is described by a linear dependence for the second and fourth order terms,
\begin{equation}
    \begin{split}
       & U^{(2)}\approx U^{(2)}_0+\alpha^{(2)}\omega \\
       & U^{(4)}\approx U^{(4)}_0+\alpha^{(4)}\omega,
    \end{split}
    \label{EQ_Somega}
\end{equation}
but that higher order coefficients are constant, $U^{(n_c\geq 6)}\approx U^{(n_c\geq 6)}_0$, where $n_c$ is the order of the interaction. This is adopted here for illustrative purposes, but our \textit{ab initio} results presented in sections \ref{bccFe} and \ref{Mn3AN} are obtained without making any assumption in this regard. Minimizing Eq.\ (\ref{EQ_Gusimp}) with respect to $\omega$ then gives
\begin{equation}
\omega=\frac{1}{V_\text{PM}\gamma}\left(
\alpha^{(2)} m^2+\alpha^{(4)} m^4
\right)
\label{EQ_w}
\end{equation}
which introduced back into Eq.\ (\ref{EQ_Gusimp}) finally provides
\begin{equation}
\begin{split}
 \mathcal{G}_\text{tot}
  &    =   -Nk_\text{B}T\ln 4\pi + U^{(0)}(V_\text{PM})
\\
 &  -\left(U^{(2)}_0-\frac{3}{2}k_\text{B} TN\right) m^2
    \\
   & -\left(U^{(4)}_0+\frac{\left[\alpha^{(2)}\right]^2}{2V_\text{PM}\gamma}-\frac{9}{20}k_\text{B} TN\right)m^4
   +\mathcal{O}(m^6),
    \label{EQ_Gusimp2}
\end{split}
\end{equation}
where $N$ is the number of magnetic atoms. To obtain the previous equation we have used the following expansion of the magnetic entropy
\begin{equation}
    S_\text{mag}=k_\text{B}\sum_n \left(
    \ln 4\pi -\frac{3}{2}m_n^2-\frac{9}{20}m_n^4 -\cdots
    \right),
    \label{EQ_Snexp}
\end{equation}
which can be derived by combining Eq.\ (\ref{EQ_Smag}) together with Eq.\ (\ref{EQ_mn}) and Eq.\ (\ref{EQ_Sn}).

%%%%%%%%%%%%%%%%%%%%%%%FIGURE%%%%%%%%%%%%%%%%%%%%%%%%%%%%%%%%%%%%%
\begin{figure}[t]
\centering
\includegraphics[clip,scale=0.6]{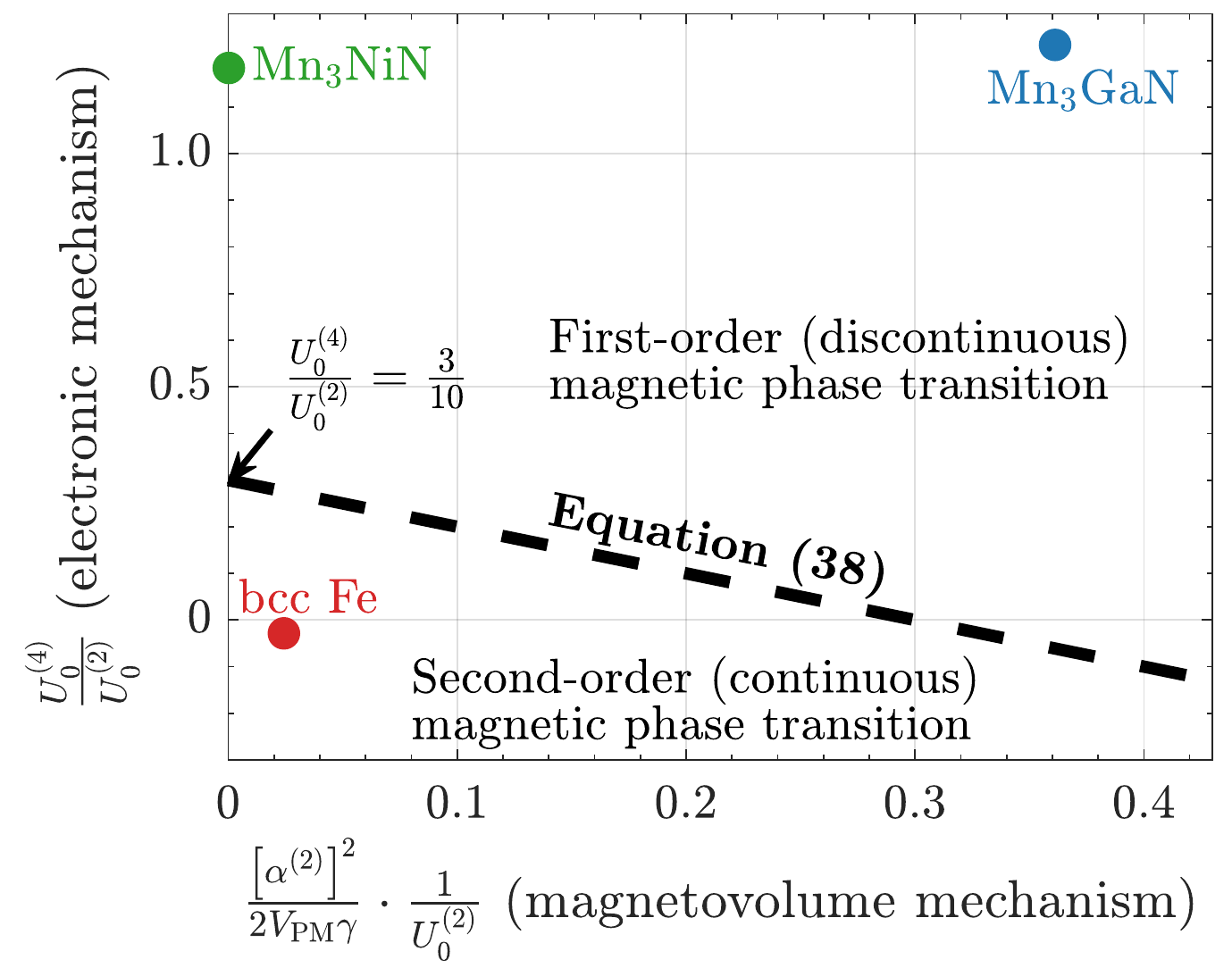}
\caption{
Critical line separating regions in which the magnetic phase transition from the paramagnetic state is second-order or first-order, as given by Eq.\ (\ref{EQ_cFOT2}). This critical line is plotted as a function of the electronic mechanism, described by the multisite term $U^{(4)}_0$, and the magnetovolume coupling, described by $\alpha^{(2)}$.
The figure shows results computed for bcc Fe (section \ref{bccFe}), as well as for Mn$_3$GaN and Mn$_3$NiN after considering a lattice thermal expansion (see section \ref{Mn3AN}).
}%
\label{FigFOTSOC}
\end{figure}
%%%%%%%%%%%%%%%%%%%%%%%%%%%%%%%%%%%%%%%%%%%%%%%%%%%%%%%%%%%%%%%%%%%%

A mathematical analysis of Eq.\ (\ref{EQ_Gusimp2}) shows that a second-order (continuous) magnetic phase transition from the paramagnetic state, i.e.,\ from $m=0$ to $m\neq 0$ by lowering the temperature, occurs when
%$\frac{\partial\mathcal{G}_\text{tot}}{\partial m}|_{m=0}=0$
$\frac{\partial^2\mathcal{G}_\text{tot}}{\partial m ^2}|_{m=0}=0$,
which gives
\begin{equation}
    T_\text{tr,sec}=\frac{ 2U^{(2)}_0 }{3k_\text{B}N},
    \label{EQ_TSOT}
\end{equation}
and that such a transition becomes discontinuous when
\begin{equation}
    \lim_{m\rightarrow 0}\frac{\partial^2\mathcal{G}_\text{tot}}{\partial m^2}\Bigg|_{T=T_\text{tr,sec}} = 0^{-}.
    \label{EQ_cFOT}
\end{equation}
Eq.\ (\ref{EQ_cFOT}) means that a first-order character arises when the overall fourth order coefficient at $T=T_\text{tr,sec}$ is negative, giving rise to the following condition
\begin{equation}
    U^{(4)}_0+\frac{\left[\alpha^{(2)}\right]^2}{2V_\text{PM}\gamma}>\frac{3}{10}U^{(2)}_0,
    \label{EQ_cFOT2}
\end{equation}
which we plot in Fig.\ \ref{FigFOTSOC}.
Eq.\ (\ref{EQ_cFOT2}) shows that a magnetovolume coupling $\alpha^{(2)}$ always contributes to enhance the first-order character of the magnetic phase transition regardless of its sign. On the other hand, $U^{(4)}_0$ has to be positive to do so, and at least 30\% of the magnitude of $U^{(2)}_0$ if the discontinuous character has a purely electronic origin. Both interactions are possible mechanisms driving magnetic phase transitions of different types and between different magnetic states. In section \ref{Mn3AN} we study the Mn-based antiperovskite materials class, which exhibits a first-order phase transition to a triangular antiferromagnetic state whose origin is mainly electronic or magnetoelastic depending on the chemical composition of the material.

%%%%%%%%%%%%%%%%%%%%%%%%%%%%%%%%%%%%%%%%%%%%%%%%%%%%%%%%%%%
\subsection{Computational details}
\label{Comp}
%%%%%%%%%%%%%%%%%%%%%%%%%%%%%%%%%%%%%%%%%%%%%%%%%%%%%%%%%%%

All our fully non-collinear DFT calculations have been performed using the Vienna Ab Initio Simulation package (VASP)~\cite{PhysRevB.47.558,PhysRevB.49.14251,PhysRevB.54.11169,KRESSE199615}. VASP already includes a method to magnetically constrain supercells to different local moment orientations $\{\hat{\textbf{e}}_n\}$, developed by Ma and Dudarev~\cite{PhysRevB.91.054420}. Albeit we use such an implementation in this work, our approach can be applied together with any other DFT code suitable to perform magnetic constraints.
Spin-orbit effects have not been considered. The parameter $\lambda$ (referred to as a Lagrange multiplier in Ref.~\cite{PhysRevB.91.054420}) describing the energy penalty term $E_p$ to the total energy functional has been increased from low to high values as $\lambda=1,3,5,7,10,15,20,25,30$ eV to preserve numerical stability within the DFT calculation. We have found that the final value of $\lambda=30$ eV provides a satisfactory energy convergence with respect to the energy scale of the disordered and partially disordered magnetic states studied. We have constrained the orientation of the local moments only so their sizes have been allowed to relax within the self-consistent calculation for each magnetic configuration $\{\hat{\textbf{e}}_n\}$, which is a fundamental feature of the time-scale separation of our theory (see section \ref{Framework}).
The Wigner-Seitz radius needed as input for magnetic constraints was 1.323\AA, 0.741\AA, 1.217\AA, and 1.286\AA\, for Mn, N, Ga, and Ni, respectively.

We have employed the projector augmented-wave (PAW) method~\cite{PhysRevB.50.17953} implemented in VASP within the Perdew–Burke–Ernzerhof (PBE) generalized gradient approximation~\cite{PhysRevLett.77.3865} to describe exchange and correlation effects.  The Brillouin zone has been sampled using a $2\times 2\times 2$ Monkhorst-Pack $k$-grid for supercells containing 432 atoms for bcc Fe ($6\times 6\times 6$ supercell of a cell of two atoms) and 320 atoms for Mn$_3$AN (A = Ga, Ni) ($4\times 4\times 4$ supercell of a cell of five atoms). The cutoff energy has been 400 eV and 500 eV, respectively. Results for these materials are shown in sections \ref{bccFe} and \ref{Mn3AN}.
In appendix \ref{App1} we show results obtained for smaller supercells.
The width of the smearing determining the partial orbital occupancies has been 0.1 eV. 
Finally, all the performed magnetically constrained DFT calculations have been supported, accelerated, and managed by Pyiron, an integrated development environment~\cite{JANSSEN201924}.

%%%%%%%%%%%%%%%%%%%%%%%FIGURE%%%%%%%%%%%%%%%%%%%%%%%%%%%%%%%%%%%%%
\begin{figure*}[t]
\centering
\includegraphics[clip,scale=0.49]{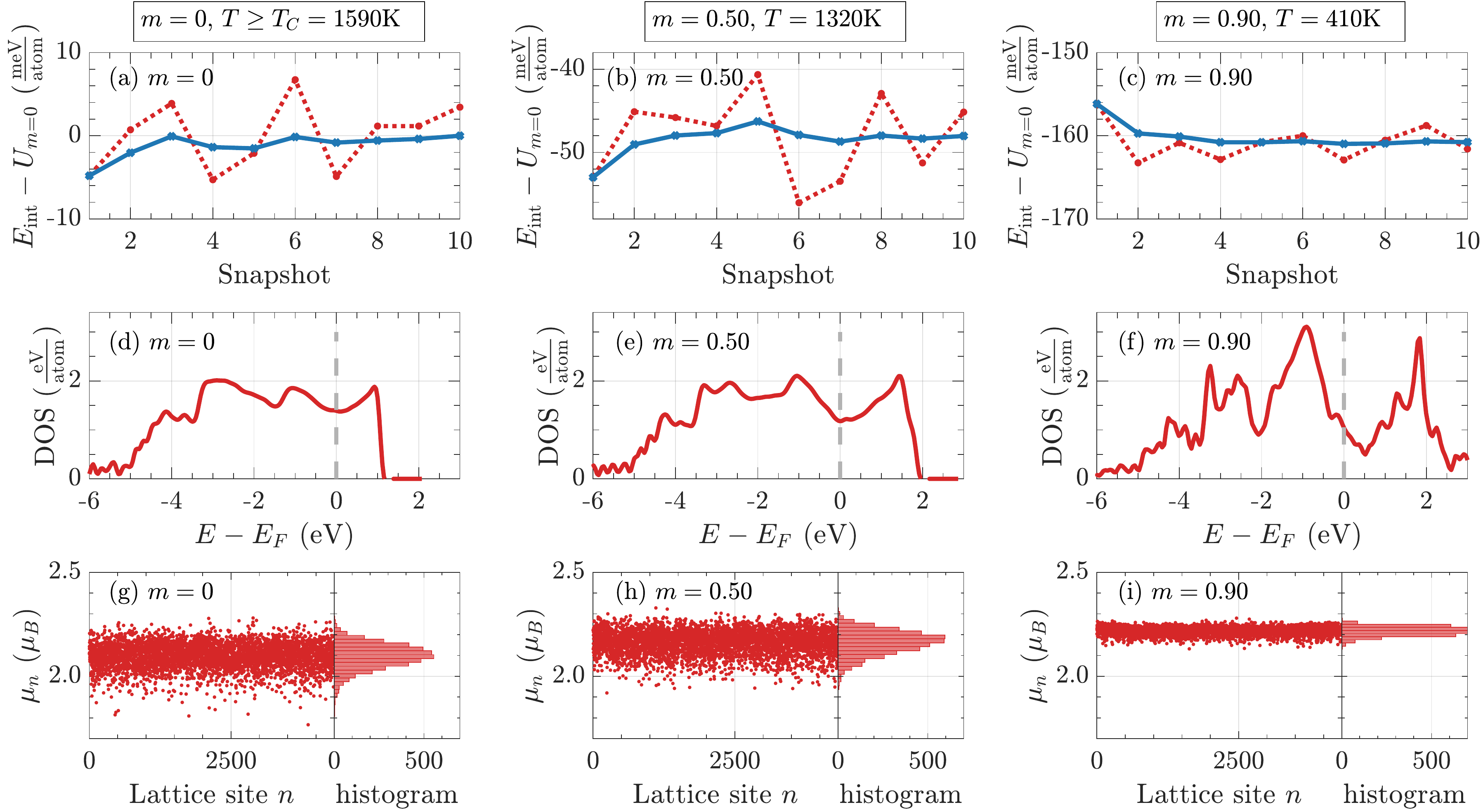}
\caption{
Magnetic and electronic properties of bcc Fe for three different thermally fluctuating states of ferromagnetic order, $m=0$ ($T\geq T_c=1590$K), $m=0.50$ ($T=1320$K), and  $m=0.90$ ($T=410$K). The associated values of the temperature are obtained only after minimizing the free energy in section \ref{bccFe2}.
(a,b,c) Total energy per atom with respect to $U_{m=0}\equiv \langle E_\text{int}\rangle_\text{tr}(m=0)$ of a $6\times 6\times 6$ supercell of a two atom cell of bcc Fe (432 atoms). Each panel shows 10 values of the energy obtained for different local moment orientation configurations $\{\hat{\textbf{e}}_n\}_i^{\{P_n(\hat{\textbf{e}}_n)\}}$ following the probability distribution $\{P_n(\hat{\textbf{e}}_n)\}$, which is prescribed by the value of $m$ (red dotted lines). The accumulative average of this quantity is also plotted (continuous blue line), which gives $U_\text{mag}=\langle E_\text{int}\rangle_\text{tr}$ through Eq.\ (\ref{EQ_AvEint_mc}).
(d,e,f) Total density of states per atom. Vertical dashed lines indicate the position of the Fermi energy.
(g,h,i) Magnitudes of the local moments, $\{\mu_n\}$, for all the sites and all 10 different magnetic configurations studied. $\{\mu_n\}$ are not constrained and so these are obtained self-consistently throughout the DFT cycle. Corresponding histograms are also shown.
}%
\label{Fig2}
\end{figure*}
%%%%%%%%%%%%%%%%%%%%%%%%%%%%%%%%%%%%%%%%%%%%%%%%%%%%%%%%%%%%%%%%%%%%

%%%%%%%%%%%%%%%%%%%%%%%%%%%%%%%%%%%%%%%%%%%%%%%%%%%%%%%%%%%
\subsection{Gibbs free energy of ferromagnetic bcc Fe}
\label{bccFe}
%%%%%%%%%%%%%%%%%%%%%%%%%%%%%%%%%%%%%%%%%%%%%%%%%%%%%%%%%%%

\subsubsection{Dependence on ferromagnetic disorder}
\label{bccFe1}

To illustrate how the magnetic Gibbs free energy of a material is obtained and minimized using our supercell-based \textit{ab initio} approach, first we present its application to the referent bcc Fe. Focusing on the study of the ferromagnetic state, it follows that all the lattice sites of bcc Fe are equivalent. This means that in our mean field theory all the local magnetic moment orientations $\{\hat{\textbf{e}}_n\}$ fluctuate with the same single-site probability distribution. The thermal fluctuations of a ferromagnetic state pointing along the $\hat{\textbf{z}}$-direction are, therefore, described by $\{\beta\textbf{h}_n=\beta h\hat{\textbf{z}}\}$, or by $\{\textbf{m}_n= m\hat{\textbf{z}}\}$. Note that $m=\frac{-1}{\beta h}+\coth(\beta h)$ from Eq.\ (\ref{EQ_mn}). The values of the single-site magnetic fields and local order parameters are consequently the same for all sites.

We first compute the dependence of the internal magnetic energy on $m$ as summarized in Table \ref{tab:scheme}. Panels (a), (b), and (c) in Fig.\ \ref{Fig2} show the total energies of 10 different orientationally constrained snapshots in $6\times 6\times 6$ supercells of 2-atom cells of bcc Fe obtained for $m=0$, $m=0.50$, and $m=0.90$, respectively (red dotted lines).
Pictures of  magnetic configurations for these values of $m$ are represented in Fig.\ \ref{Fig1a}.
The lattice parameter used is $a=2.83$\AA, which is the value that minimizes the energy in the ferromagnetic state within our DFT approximations~\cite{PhysRevB.87.214102}.
We have found that the average of the total energy over $N_\text{MC}=10$ different magnetic configurations for each value of $m$ suffices to compute $U_\text{mag}=\langle E_\text{int}\rangle_\text{tr}$ using Eq.\ (\ref{EQ_AvEint_mc}) within an error of only a small number of meV per atom (continuous blue line).
The energy scale of the magnetic thermal fluctuations for the ferromagnetic state of bcc Fe is such that $\langle E_\text{int}\rangle_\text{tr}(m=0)-\langle E_\text{int}\rangle_\text{tr}(m=1)\approx 200$meV/atom. We, therefore, can conclude that the application of Eq.\ (\ref{EQ_AvEint_mc}) provides a satisfactory accuracy to obtain the dependence of the internal magnetic energy on sizable changes of $m$. However, the description of quantities with a smaller energy scale, such as the magnetocrystalline anisotropy, would require one to increase $N_\text{MC}$ and/or the size of the supercell.

The magnetic disorder characterizing the highly fluctuating paramagnetic state ($m=0$) covers a wide range of magnetic excitations, which directly impacts the underlying electronic structure. Disorder typically broadens the spectral density function of electronic states in perfect periodic metals~\cite{PhysRevB.21.3222,PhysRevB.73.205109,PhysRevB.94.224205}. Consequently, a broadening of the total density of states (DOS) is also expected, as shown in Fig.\ \ref{Fig2}(d). Indeed, decreasing the amount of magnetic disorder by increasing the value of $m$ directly reduces the number of magnetic excitations thus sharpening the peaks of the DOS in Fig.\ (\ref{Fig2})(e,f).

\begin{figure}[t]
\centering
\includegraphics[clip,scale=0.48]{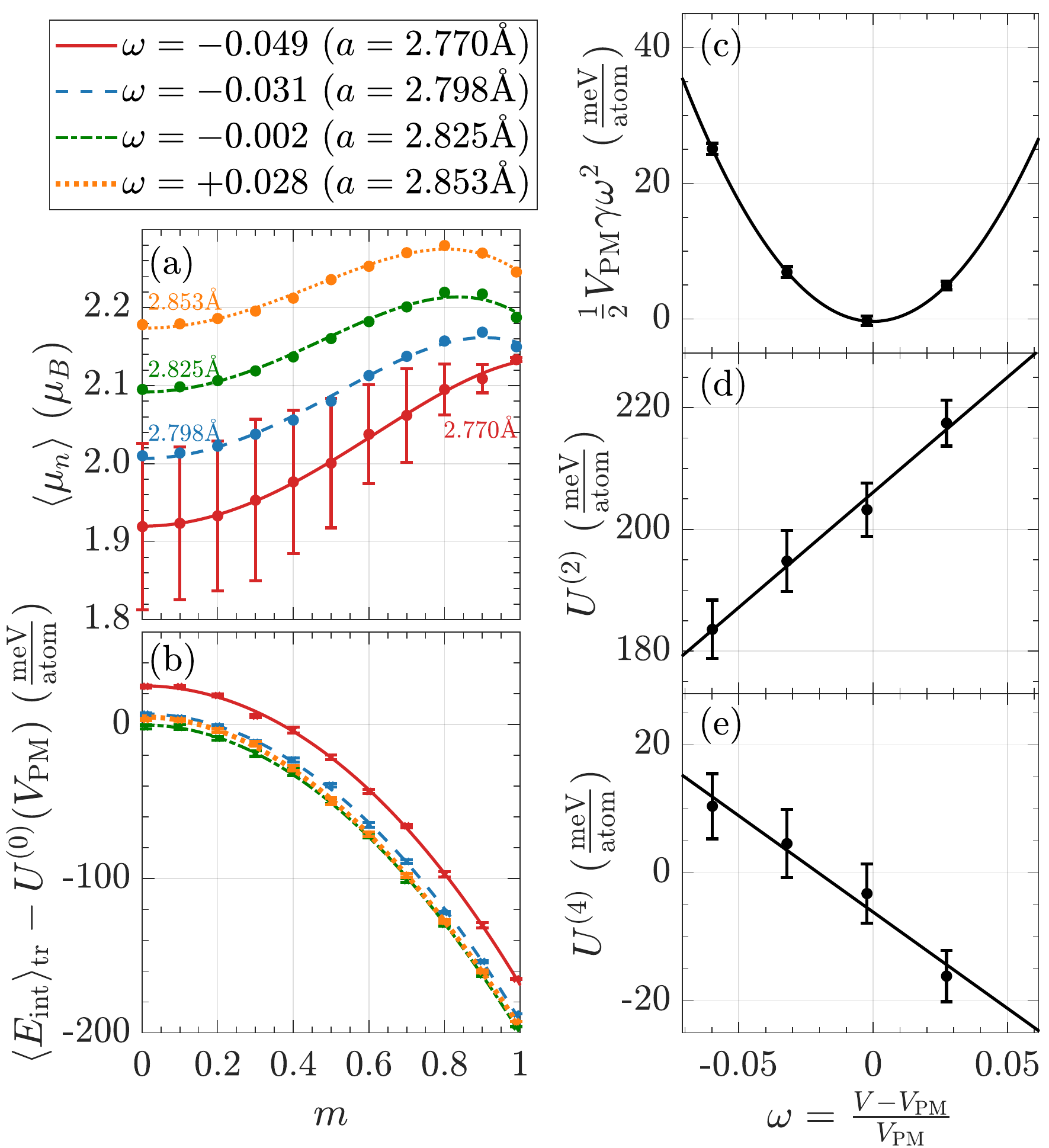}
\caption{
Magnetic properties of bcc Fe at different values of the ferromagnetic order parameter and of the lattice parameter $a$, which link to different volume changes relative to the paramagnetic state, $\omega=\frac{V-V_\text{PM}}{V_\text{PM}}$.
(a) Mean value of the local moment magnitudes against orientational ferromagnetic order $m$. Vertical bars indicate the standard deviation. Since these have very similar magnitudes for all curves, we show them only for one of them for clarity.
(b) Internal magnetic energy $U_\text{mag}=\langle E_\text{int}\rangle_\text{tr}$ per atom as a function of $m$ with respect to its value in the paramagnetic state [$U^{(0)}(V_\text{PM})$, $m=0$]. The computed \textit{ab initio} data is plotted with points, together with bars showing the standard error. Continuous lines correspond to the least-squares regression of Eq.\ (\ref{EQ_EintbccFe}).
(c,d,e) Internal energy coefficients $\{\frac{1}{2}V_\text{PM}\gamma\omega^2,U^{(2)},U^{(4)}\}$ against $\omega$ (data points), where $U^{(0)}-U^{(0)}(V_\text{PM})=\frac{1}{2}V_\text{PM}\gamma\omega^2$. Their linear regressions are also shown (continuous lines). All the results are given in meV/atom and error bars associated to the regressions are provided.	
}%
\label{Fig3}
\end{figure}

The time-scale separation between slowly evolving local moment orientations $\{\hat{\textbf{e}}_n\}$ and a rapidly adapting electronic structure is a central tenet of our DLM theory. Within this framework we allow the magnitudes of the local moments $\{\mu_n\}$ to relax for a given magnetic configuration of the orientations within the self-consistent DFT cycle. In other words, we assume that these longitudinal fluctuations are much faster than the transverse ones and that their energy scale is much larger. In Fig.\ \ref{Fig2}(g,h,i) we show the relaxed values of $\{\mu_n\}$ for all the sites of the 10 configurations $\{\hat{\textbf{e}}_n\}$ obtained for $m=0$, $m=0.50$, and $m=0.90$. For an almost fully ordered ferromagnetic state ($m=0.90$) the sizes of the local moments vary very little around their mean value. However, for highly disordered configurations ($m=0$, $m=0.50$), $\mu_n$ can deviate substantially owing to non-ferromagnetic environments surrounding a given site $n$. The mean value of the magnetic moment magnitude, $\langle\mu_n\rangle$, decreases slightly by lowering the order parameter from $m=0.50$ to $m=0$. Interestingly, we have found a non-monotonous behavior around $m=0.80$ at which a maximum of $\langle\mu_n\rangle$ occurs. This is shown in Fig.\ \ref{Fig3}(a), where we plot $\langle\mu_n\rangle$ against $m$ for different volumes. The vertical bars indicate the standard deviation, which becomes larger by decreasing $m$ in accordance to Fig.\ \ref{Fig2}(g,h,i).
To obtain the temperature dependence, which is encoded in $m(T)$, we need to minimize the free energy with respect to $m$ at different values of $T$, which we address in the following.

%%%%%%%%%%%%%%%%%%%%%%%FIGURE%%%%%%%%%%%%%%%%%%%%%%%%%%%%%%%%%%%%%
%%%%%%%%%%%%%%%%%%%%%%%%%%%%%%%%%%%%%%%%%%%%%%%%%%%%%%%%%%%%%%%%%%%%

\subsubsection{Minimization of the free energy at different temperatures}
\label{bccFe2}

The construction of the free energy, $\mathcal{G}_\text{tot}$, requires one to firstly obtain $U_\text{mag}=\langle E_\text{int}\rangle_\text{tr}$ for different values of $m$, which must follow a polynomial dependence of even terms
[see Eq.\ (\ref{EQ_Eintmn})]. In Fig.\ \ref{Fig3}(b) we show $\langle E_\text{int}\rangle_\text{tr}$ against $m$ computed for different volumes $V$, which we can express in terms of $\omega$ (see section \ref{MultiMVE}).
We have found that the \textit{ab initio} data obtained can be described very well by
\begin{equation}
    \langle E_\text{int}\rangle_\text{tr}-U^{(0)}(V_\text{PM})=
    \frac{1}{2}V_\text{PM}\gamma\omega^2
    -U^{(2)}m^2
    -U^{(4)}m^4,
    \label{EQ_EintbccFe}
\end{equation}
where we have subtracted the value of the paramagnetic energy at $V_\text{PM}$, $U^{(0)}(V_\text{PM})$, such that the first term in the right hand side becomes $U^{(0)}-U^{(0)}_{V_\text{PM}}=\frac{1}{2}V_\text{PM}\gamma\omega^2$ [see Eq.\ (\ref{Eq_E0vare})].
A least-squares regression of Eq.\ (\ref{EQ_EintbccFe}) can be performed to obtain $\{\frac{1}{2}V_\text{PM}\gamma\omega^2,U^{(2)},U^{(4)}\}$ for the different volumes studied. We show the results in Fig.\ \ref{Fig3}(c,d,e) with data points. 
Fig.\ \ref{Fig3}(c) demonstrates that the internal magnetic energy in the paramagnetic state, $U^{(0)}$, follows a pronounced parabolic behavior with a minimum at $V_\text{PM}$, as stated in Eq.\ (\ref{Eq_E0vare}). On the other hand, $U^{(2)}$ and $U^{(4)}$ show a linear dependence, following the behavior in Eq.\ (\ref{EQ_Somega}).

{\renewcommand{\arraystretch}{1.5}
\begin{table}[t]
    \centering
    \begin{tabular}{|cc|cc|cc|}
    \hline
      $V_\text{PM}$ &  $\gamma$ &  $U^{(2)}_0$    & $\alpha^{(2)}$  & $U^{(4)}_0$  & $\alpha^{(4)}$  \\ 
     (\AA$^3$) & (GPa) &  $\left(\frac{\text{meV}}{\text{atom}}\right)$  & $\left(\frac{\text{meV}}{\text{atom}}\right)$  & $\left(\frac{\text{meV}}{\text{atom}}\right)$  & $\left(\frac{\text{meV}}{\text{atom}}\right)$  \\   \hline
    11.30$\pm$0.09 & 202$\pm$2 & 206$\pm$1 & 378$\pm$26 & -6.1$\pm$1.3 & -300$\pm$40 \\
    \hline
    \end{tabular}
    \caption{Coefficients describing the internal magnetic energy, the bulk modulus in the paramagnetic limit ($\gamma$), and the magnetovolume coupling. These are obtained by performing a least-squares regression of $U^{(0)}(V)=U^{(0)}(V_\text{PM})+\frac{1}{2}V_\text{PM}\gamma\omega^2$ and of Eq.\ (\ref{EQ_Somega}). The value of $V_\text{PM}$ given corresponds to a unit cell containing a single atom. Error estimations associated to the regression are also given.}
    \label{Table1}
\end{table}
}

We can perform another regression, now of $\{U^{(0)},U^{(2)},U^{(4)}\}$ with respect to $\omega$, to obtain $\{V_\text{PM},\gamma,U^{(2)}_0,\alpha^{(2)},U^{(4)}_0,\alpha^{(4)}\}$ using Eqs.\ (\ref{Eq_E0vare}) and (\ref{EQ_Somega}). We show the results obtained in Table \ref{Table1}.
The bulk modulus in the paramagnetic state that we have computed agrees satisfactorily with experiment~\cite{PhysRev.122.1714}.
Most importantly, we have found that $\frac{U^{(4)}_0}{U^{(2)}_0}=-0.03\ll \frac{3}{10}$ and $\frac{\left(\alpha^{(2)}\right)^2}{2V_\text{PM}\gamma U^{(2)}_0}=0.02\ll \frac{3}{10}$. According to Eq.\ (\ref{EQ_cFOT2}), these numbers directly imply that the fourth order terms in the free energy coming from the purely electronic source $U^{(4)}_0$ or from the magnetovolume coupling $\alpha^{(2)}$ cannot generate a first-order character for the transition. See Fig.\ \ref{FigFOTSOC} for a graphical representation of the position of bcc Fe with respect to the critical line defining the condition of a first-order transition. This means that the paramagnetic-ferromagnetic phase transition is continuous and occurs at $T_c=\frac{2U^{(2)}_0}{3k_\text{B}N}=1590$K as given by Eq.\ (\ref{EQ_TSOT}) (note that the value of $U^{(2)}_0$ in Table \ref{Table1} is already given in units of energy per atom). The computed value of $T_c$ is somewhat larger than the experimental one~\cite{doi:10.1098/rspa.1955.0102}, but in satisfactory agreement considering the mean-field nature of our theory. A similar value has also been obtained using a mean-field and random phase approximation~\cite{PhysRevB.78.033102}. We also note that the leading magnetovolume coefficient, $\alpha^{(2)}$, is fairly small, which also agrees with experiment~\cite{doi:10.1098/rspa.1955.0102}.

%%%%%%%%%%%%%%%%%%%%%%%FIGURE%%%%%%%%%%%%%%%%%%%%%%%%%%%%%%%%%%%%%
\begin{figure}[t]
\centering
\includegraphics[clip,scale=0.58]{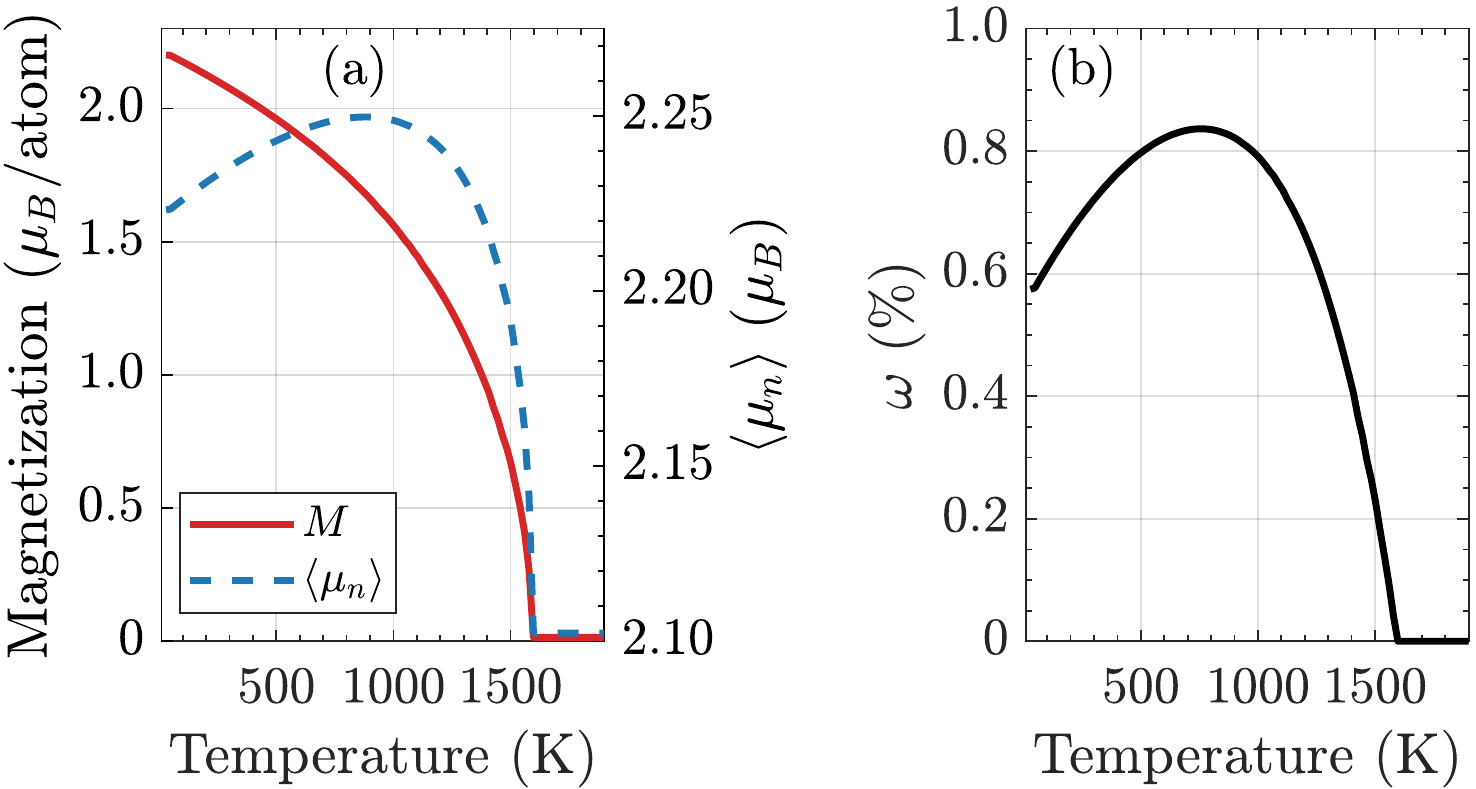}
\caption{
Dependence on temperature of (a) the total magnetization, $M$, the mean value of the magnitude of the local moments, $\langle\mu_n\rangle$, and of (b) the relative volume change, $\omega$, obtained for bcc Fe.
}%
\label{Fig4}
\end{figure}
%%%%%%%%%%%%%%%%%%%%%%%%%%%%%%%%%%%%%%%%%%%%%%%%%%%%%%%%%%%%%%%%%%%%

Introducing the values obtained for $\{V_\text{PM},\gamma,U^{(2)}_0,\alpha^{(2)},U^{(4)}_0,\alpha^{(4)}\}$ into Eq.\ (\ref{EQ_EintbccFe}) allows one to construct $\mathcal{G}_\text{tot}$ [Eq.\ (\ref{EQ_Gusimp})], whose analytical minimization with respect to $\omega$ is given by Eq.\ (\ref{EQ_w}).
We can, therefore, compute the dependence on temperature of the magnetization and other properties by numerically finding the value of $\beta h$ that minimizes Eq.\ (\ref{EQ_Gusimp}) at different values of $T$, which is equivalent to perform a minimization with respect to $m$.
Fig.\ \ref{Fig4} shows the results obtained for $\langle\mu_n\rangle(T)$ and $\omega(T)$, as well as for the total magnetization per atom, $M(T)=m(T)\cdot\langle\mu_n\rangle(T)$. Indeed, the paramagnetic-ferromagnetic phase transition is continuous and occurs at $T_c=1590$K.
Albeit $\langle\mu_n\rangle(T)$ presents a non-monotonic dependence on $T$, it is negligible and so the total magnetization behaves monotonously because its major contribution comes from $m(T)$.
A prediction of our calculations is that $\alpha^{(2)}$ is positive, which directly implies that the volume  increases by lowering the temperature through $T_c$, the so-called negative volume expansion (NVE). However, the computed effect is very small in comparison with the intrinsic lattice thermal expansion measured experimentally~\cite{doi:10.1098/rspa.1955.0102}. On the other hand, $\alpha^{(4)}$ is negative and sufficiently large to cause a decrease of $\omega$ at lower temperatures when the fourth order free energy terms become important.

%%%%%%%%%%%%%%%%%%%%%%%%%%%%%%%%%%%%%%%%%%%%%%%%%%%%%%%%%%%
%%%%%%%%%%%%%%%%%%%%%%%%%%%%%%%%%%%%%%%%%%%%%%%%%%%%%%%%%%%
\section{\textit{Ab initio} origin of first-order magnetic phase transitions in antiperovskite nitride materials}
\label{Mn3AN}
%%%%%%%%%%%%%%%%%%%%%%%%%%%%%%%%%%%%%%%%%%%%%%%%%%%%%%%%%%%
%%%%%%%%%%%%%%%%%%%%%%%%%%%%%%%%%%%%%%%%%%%%%%%%%%%%%%%%%%%

Mn-based antiperovskite nitride systems Mn$_3$AN, where A can be one or a solution among different transition metals and semiconductor elements, form a magnetic materials class that has been largely studied over the last decades owing to their interesting magnetic properties~\cite{FRUCHART19711793,doi:10.1143/JPSJ.44.781,doi:10.1063/1.2147726,doi:10.1088/1468-6996/15/1/015009,PhysRevB.96.024451,PhysRevB.99.104428,PhysRevMaterials.3.094409,PhysRevMaterials.4.024408,doi:10.1021/acs.chemmater.8b01618,Singh2021,doi:10.7566/JPSJ.90.044601}. 
The corresponding cubic crystal structure belongs to the perovskite space group 221 in which A, N, and Mn atoms sit on corners, centers, and face centers of the unit cell, respectively.
In these materials a magnetic phase transition from the paramagnetic state to a triangular antiferromagnetic state occurs at different temperatures and emerges from the geometrical frustration of antiferromagnetic interactions between nearest neighbor Mn atoms.
The transition is accompanied by a substantial NVE~\cite{LB1981,doi:10.1063/1.2147726,doi:10.1088/1468-6996/15/1/015009} and has a first-order (discontinuous) character, which provides a giant barocaloric effect with great potential in the field of caloric refrigeration~\cite{Matsunami1,TAO2021114049,PhysRevB.104.134101}.

We have recently made an analysis of Mn$_3$AN materials based on the exploration of free energy coefficients in Eq.\ (\ref{EQ_Gusimp}) that reproduce experimental data describing the change of the magnetic phase transition under applied hydrostatic pressure. Such an experimentally-guided study has shown that the first-order character of Mn$_3$AN arises from the combined effect of the magnetovolume coupling and multisite magnetic interactions, and that the choice of atom A selects which one of these sources is the predominant mechanism~\cite{PhysRevX.8.041035}. Here we use our supercell DLM approach to provide a full frist-principles demonstration of this phenomenon.

We focus on the study of Mn$_3$GaN and Mn$_3$NiN, whose first-order character should be mainly driven by a magnetovolume coupling and multisite interactions, respectively~\cite{PhysRevX.8.041035}.
The triangular antiferromagnetic state that stabilizes in these materials is a non-modulated ($q=0$) magnetic structure. Here the three Mn magnetic local moments lie within the same plane and form relative angles of 120 degrees at $T=0$K when magnetic anisotropy is not considered. If the cubic crystal structure is not distorted, such a triangular state is fully compensated, i.e.\ there is no overall magnetization. This symmetry also means that the thermal fluctuations of the local moment orientations at the three sites can be described by triangular local order parameters that have the same magnitude $m_\text{tri}$. 
Since we do not consider spin-orbit effects we simply choose an arbitrary plane and apply Eq.\ (\ref{EQ_theta}) and Eq.\ (\ref{EQ_phi}) to find the polar and azimuthal angles for each moment in a supercell. The local axis frames are rotated accordingly to the triangular antiferromagnetic state such that $\{\textbf{h}_n\}$ (or $\{\textbf{m}_n\}$) correctly form angles of 120 degrees between the three non-equivalent magnetic sub-lattices.

%%%%%%%%%%%%%%%%%%%%%%%FIGURE%%%%%%%%%%%%%%%%%%%%%%%%%%%%%%%%%%%%%
\begin{figure}[t]
\centering
\includegraphics[clip,scale=0.48]{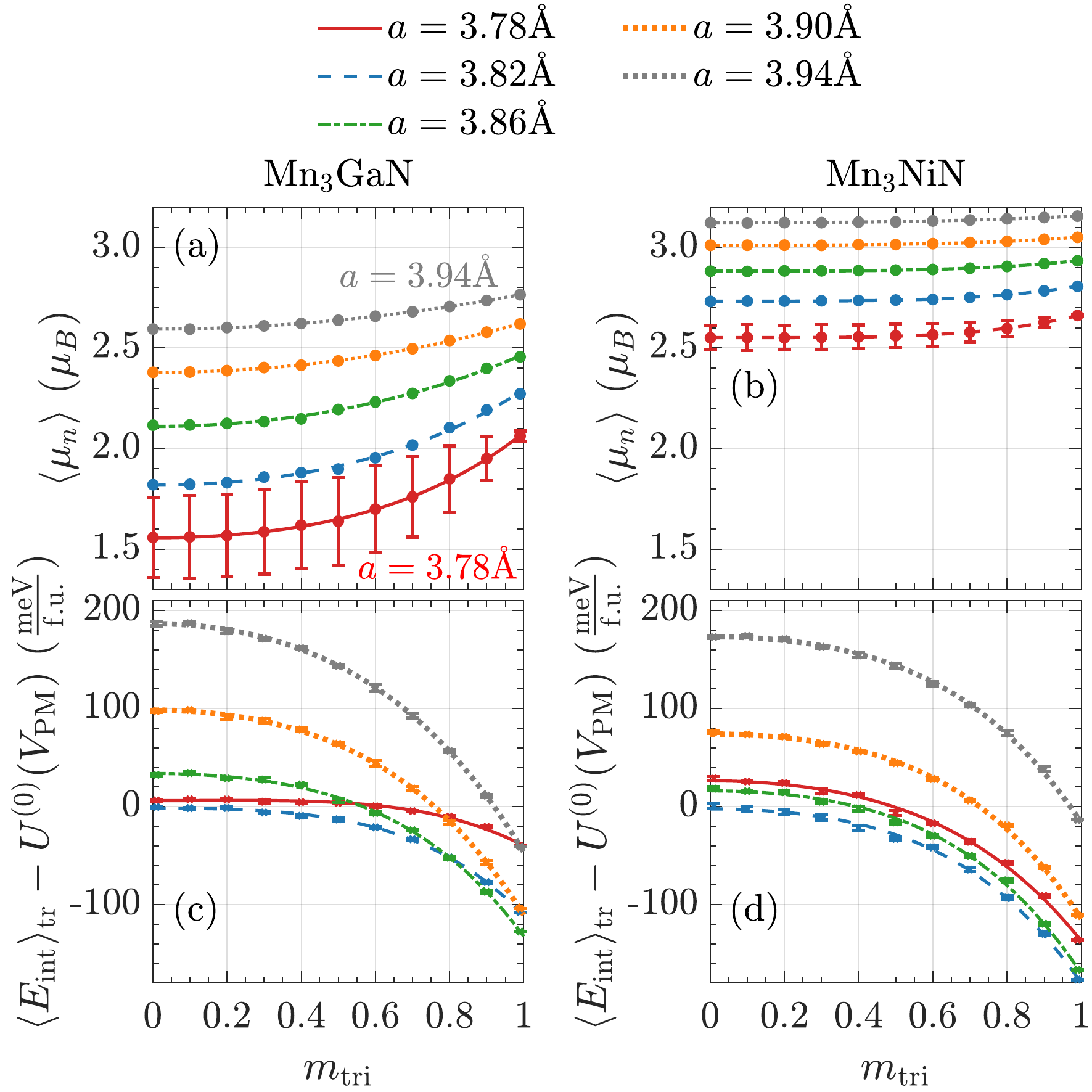}
\caption{
Dependence on the amount of triangular antiferromagnetic order, $m_\text{tri}$, of the mean local moment magnitude in Mn atoms [panels (a) and (b)] and of the internal magnetic energy  [panels (c) and (d)] computed for Mn$_3$GaN and Mn$_3$NiN  at different volumes.
The vertical bars in panels (a) and (b) correspond to the standard deviation, which is shown only for one curve for clarity.
Bars associated to the standard error of the internal energy average in panels (c) and (d) are also shown, as well as curves corresponding to their regressions.
}%
\label{Fig5}
\end{figure}
%%%%%%%%%%%%%%%%%%%%%%%%%%%%%%%%%%%%%%%%%%%%%%%%%%%%%%%%%%%%%%%%%%%%

In Fig.\ \ref{Fig5} we show the dependence of $\langle\mu_n\rangle$ and the internal magnetic energy on $m_\text{tri}$ for different volumes.
One can see in panels (a) and (b) that the local moment sizes in Mn atoms are statistically smaller and have a larger standard deviation in Mn$_3$GaN than those computed in Mn$_3$NiN. However, robust local moments emerge in both cases for larger volumes. We note that some of our calculations presented difficulties to converge for Mn$_3$GaN at $a=3.78$\AA\, and $m<0.2$, for which the local moment magnitudes are relatively small. Contrary to the ferromagnetic state of bcc Fe in section \ref{bccFe}, here $\langle\mu_n\rangle$ follows a full monotonic behavior with $m$ at all studied volumes. Very different outcomes have been obtained regarding the calculation of $U_\text{mag}=\langle E_\text{int}\rangle_\text{tr}$: Its curvature found in Mn$_3$NiN is fairly insensitive to the volume, which is in sharp contrast to Mn$_3$GaN. This means that Mn$_3$GaN and Mn$_3$NiN contain comparatively large and small magnetovolume coupling, respectively, as observed experimentally~\cite{PhysRevX.8.041035}.

The dependence of $\langle E_\text{int}\rangle_\text{tr}$ on the order parameter can be described rather well by an expansion up to a fourth order,
\begin{equation}
    \langle E_\text{int}\rangle_\text{tr}-U^{(0)}(V_\text{PM})=
    \frac{1}{2}V_\text{PM}\gamma\omega^2
    -U^{(2)}m_\text{tri}^2
    -U^{(4)}m_\text{tri}^4.
    \label{EQ_EintMn3AN}
\end{equation}
We can carry out again two consecutive least-squares regressions. The first for Eq.\ (\ref{EQ_EintMn3AN}) to obtain the corresponding internal energy coefficients, and the second to fit these against the different volumes studied.

%%%%%%%%%%%%%%%%%%%%%%%FIGURE%%%%%%%%%%%%%%%%%%%%%%%%%%%%%%%%%%%%%
\begin{figure}[t]
\centering
\includegraphics[clip,scale=0.54]{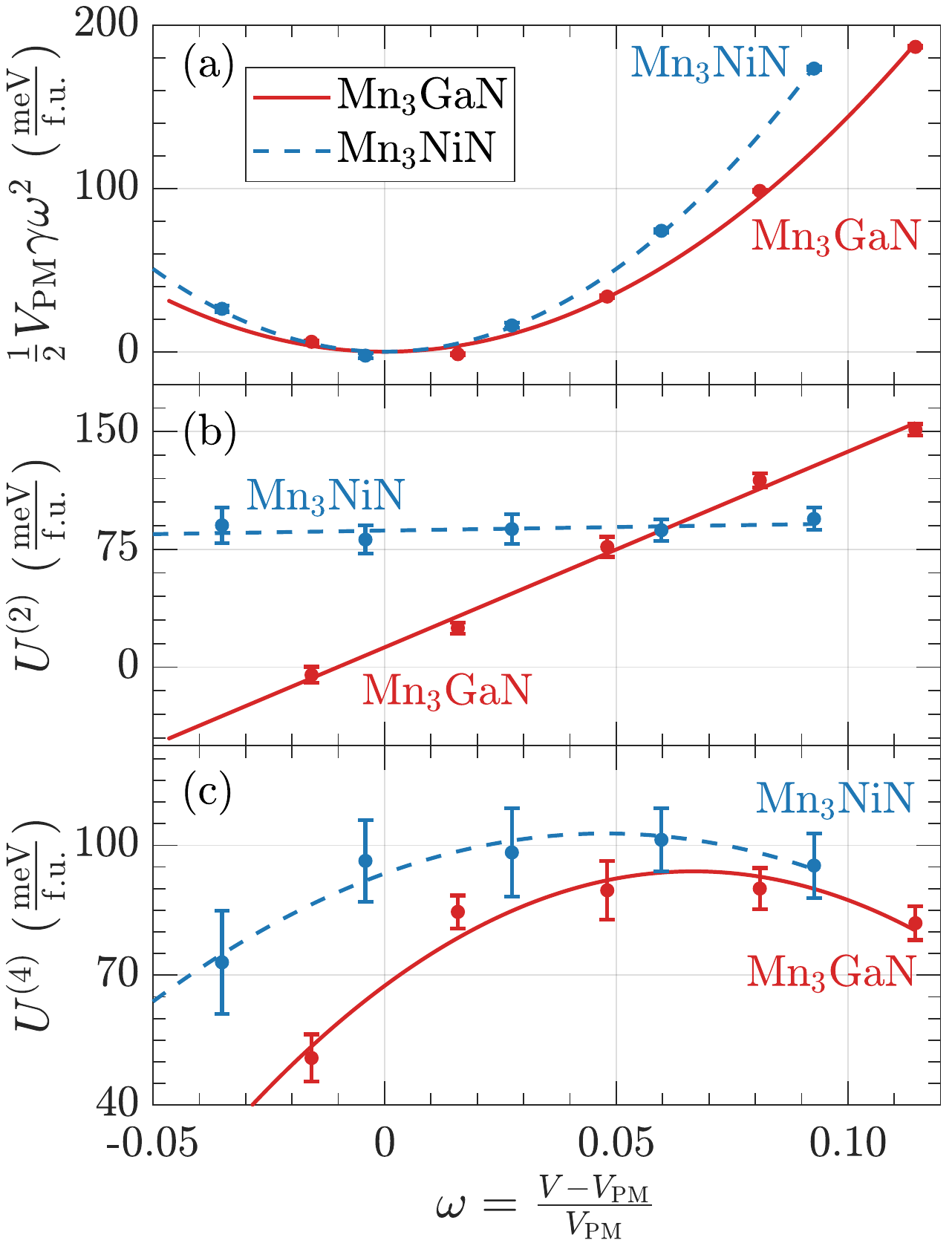}
\caption{
Dependence on the relative volume change $\omega$ of the free energy coefficients describing the internal magnetic energy of Mn$_3$GaN and Mn$_3$NiN. Data points and associated error bars correspond to the values obtained from a least-squares regression of Eq.\ (\ref{EQ_EintMn3AN}), and continuous lines are the outcome given by another regression of Eq.\ (\ref{EQ_SomegaMn3AN}).
}%
\label{Fig6}
\end{figure}
%%%%%%%%%%%%%%%%%%%%%%%%%%%%%%%%%%%%%%%%%%%%%%%%%%%%%%%%%%%%%%%%%%%%

We show the results obtained from the first regression in Fig.\ \ref{Fig6} as data points. An inspection of the figure reveals that in both materials $U^{(0)}$ and $U^{(4)}$ follow parabolic behaviors. However, we have found a distinctive feature regarding $U^{(2)}$, which is fairly constant in Mn$_3$NiN while in Mn$_3$GaN it shows a substantial linear dependence. We thus write
\begin{equation}
    \begin{split}
       & U^{(0)}=U^{(0)}(V_\text{PM})+\frac{1}{2}\gamma V_\text{PM}\omega^2, \\
       & U^{(2)}= U^{(2)}_0+\alpha^{(2)}\omega, \\
       & U^{(4)}= U^{(4)}_0+\alpha^{(4)}\omega+\beta^{(4)}\omega^2,
    \end{split}
    \label{EQ_SomegaMn3AN}
\end{equation}
which contains an additional term for a quadratic magnetovolume coupling described by $\beta^{(4)}$, in comparison to Eq.\ (\ref{EQ_Somega}). In Table \ref{Table2} we show the coefficients of Eq.\ (\ref{EQ_SomegaMn3AN}) obtained by performing the second regression with respect to $\omega$. We use continuous lines in Fig.\ \ref{Fig6} to plot the corresponding results. The bulk moduli that we have computed in the paramagnetic limit are 84~GPa and 117~GPa for Mn$_3$GaN and Mn$_3$NiN, respectively. These values are somewhat smaller compared with DFT calculations made at zero temperature~\cite{PhysRevB.96.024451}, i.e.\ magnetic thermal fluctuations reduce the bulk modulus.
Most importantly, the leading second order magnetovolume term, $\alpha^{(2)}$, is very large for Mn$_3$GaN but negligible for Mn$_3$NiN, as already seen in Fig.\ \ref{Fig5}(c,d) and reported in experiment~\cite{PhysRevX.8.041035}.

%%%%%%%%%%%%%%%%%%%%%%%%%%%%%%%%%%%%%%%%%%%%%%%%%%%%%%%%%%%%%%%%%%%%
{\renewcommand{\arraystretch}{1.7}
\begin{table}[t]
    \centering
    \begin{tabular}{|c|cc|}
    \hline
                                     & Mn$_3$GaN        & Mn$_3$NiN \\  \hline\hline
    $V_\text{PM}$ (\AA$^3$)          & 55 $\pm$ 7       & 56 $\pm$ 3  \\
    $\gamma$ (GPa)                   & 84 $\pm$ 14      & 117 $\pm$ 7 \\
    $U^{(2)}_0$ (meV/f.u.)           & 13 $\pm$ 4      & 87 $\pm$ 3  \\
    $\alpha^{(2)}$ (meV/f.u.)        & 1250 $\pm$ 60    & 45 $\pm 50$ ($\approx 0$) \\
    $U^{(4)}_0$ (meV/f.u.)           & 68 $\pm$ 4       & 93 $\pm$ 3 \\
    $\alpha^{(4)}$ (meV/f.u.)        & 793 $\pm$ 150    & 388 $\pm$ 70 \\
    $\beta^{(4)}$ (meV/f.u.)         & -5950 $\pm$ 1400  & -4070 $\pm$ 1100 \\
    $\frac{\left(\alpha^{(2)}\right)^2}{2V_\text{PM}\gamma}$ (meV/f.u.)    & 27 & 0.02 ($\approx 0$)  \\
    \hline
    \end{tabular}
    \caption{Internal energy coefficients describing the dependence on the relative volume change $\omega$ given in Eq.\ (\ref{EQ_SomegaMn3AN}) computed for Mn$_3$GaN and Mn$_3$NiN. These include the bulk modulus in the paramagnetic limit ($\gamma$) and other free energy parameters ($\{U^{(2)}_0\alpha^{(2)},U^{(4)}_0,\alpha^{(4)},\beta^{(4)}\}$), obtained by performing a regression of Eq.\ (\ref{EQ_SomegaMn3AN}).
    Errors associated to the regression are provided.}
    \label{Table2}
\end{table}
}
%%%%%%%%%%%%%%%%%%%%%%%%%%%%%%%%%%%%%%%%%%%%%%%%%%%%%%%%%%%%%%%%%%%%

%%%%%%%%%%%%%%%%%%%%%%%FIGURE%%%%%%%%%%%%%%%%%%%%%%%%%%%%%%%%%%%%%
\begin{figure}[t]
\centering
\includegraphics[clip,scale=0.56]{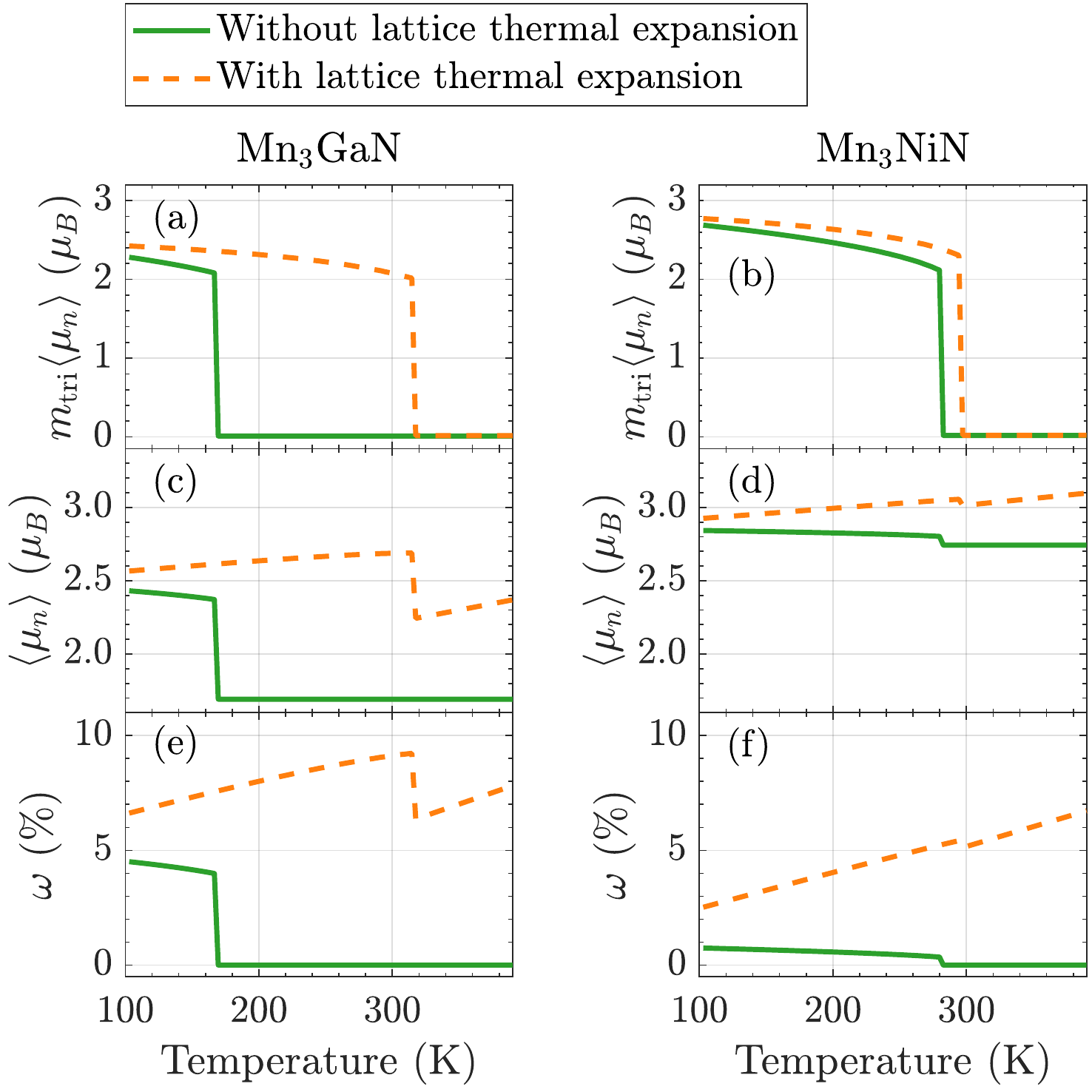}
\caption{
Temperature dependence of (a,b) the net magnetic moment per site, (c,d) mean value of the local moment magnitudes ($\langle\mu_n\rangle$), and of (e,f) the relative volume change $\omega$. The net magnetic moment per site is given by $m_\text{tri}(T)\cdot\langle\mu_n\rangle(T)$.
}%
\label{Fig7}
\end{figure}
%%%%%%%%%%%%%%%%%%%%%%%%%%%%%%%%%%%%%%%%%%%%%%%%%%%%%%%%%%%%%%%%%%%%

The free energy per formula unit of both materials is
\begin{equation}
\begin{split}
    \frac{1}{N_\text{sc}}\mathcal{G}_\text{tot}
   = & -3TS_\text{mag}+U^{(0)}(V_\text{PM})+\frac{1}{2}V_\text{PM}\gamma\omega^2 \\
   & -U^{(2)}(\omega)m_\text{tri}^2-U^{(4)}(\omega)m_\text{tri}^4,
    \label{EQ_Mn3ANG}
\end{split}
\end{equation}
where $N_\text{sc}$ is the number of Mn$_3$AN cells with volume $V_\text{PM}$ (containing five atoms) forming the supercell, and so here $\{U^{(0)},U^{(2)},U^{(4)}\}$ are also given per formula unit. The factor of 3 accompanying the entropy term comes from the fact that one crystallographic unit cell contains three Mn atoms.
Minimizing Eq.\ (\ref{EQ_Mn3ANG}) provides the total magnetization, $\langle\mu_n\rangle$, and $\omega$ as functions of $T$. An analytical expression for the latter is directly given by
\begin{equation}
\omega=\frac{1}{V_\text{PM}\gamma-2\beta^{(4)}m_\text{tri}^4}
\left(\alpha^{(2)} m_\text{tri}^2+\alpha^{(4)} m_\text{tri}^4\right).
\label{EQ_wMn3AN}
\end{equation}
In Fig.\ \ref{Fig7} we show the results obtained after performing this minimization.
The figure demonstrates that first-order magnetic phase transitions from the paramagnetic state to the triangular antiferromagnetic state are directly described by our theory and that a NVE at the transition is correctly captured. The application of Eq.\ (\ref{EQ_cFOT2}), which requires the calculation of $\left[\alpha^{(2)}\right]^2/2V_\text{PM}\gamma$  (given in Table \ref{Table2}), shows that the origin of the first-order character in Mn$_3$NiN is the purely electronic term $U^{(4)}$, as found experimentally~\cite{PhysRevX.8.041035}. On the other hand, in Mn$_3$GaN both $U^{(4)}$ and the magnetovolume coupling $\alpha^{(2)}$ are important factors behind the discontinuous character of the transition (see also Fig.\ \ref{FigFOTSOC}).

Values computed for both the transition temperature, $T_\text{tr}$, and the spontaneous volume change at the transition, $\Delta\omega$, provided in Table \ref{Table3}, agree well with experiment for Mn$_3$NiN. However, we underestimate the former and overestimate the latter for Mn$_3$GaN~\cite{PhysRevX.8.041035,Singh2021}. We point out that $T_\text{tr}$ strongly increases with $U^{(2)}_0$. Hence, according to the results given in Fig.\ \ref{Fig6}(b) a lattice thermal expansion, not considered so far but always present in materials, should substantially enhance $T_\text{tr}$ in Mn$_3$GaN while having a little impact in Mn$_3$NiN. This is indeed what we have found when we model this effect by modifying the third term in the right hand side of Eq.\ (\ref{EQ_Mn3ANG}) as
\begin{equation}
    \frac{1}{2}V_\text{PM}\gamma\omega^2
    \rightarrow
    \frac{1}{2}V_\text{PM}\gamma
    \left(\omega-C_\text{TE} T\right)^2,
    \label{EQ_ThermExp}
\end{equation}
where $C_\text{TE}>0$ drives an increment of $\omega$ when $T$ increases. Note that now Eq.\ (\ref{EQ_wMn3AN}) becomes $\omega=\omega|_{C_\text{TE}=0}+C_\text{TE} T$.
Dashed curves in Fig.\ \ref{Fig7} show how the magnetic properties change when $C_\text{TE}=40\times 10^{-4}$meV/K and $C_\text{TE}=35\times 10^{-4}$meV/K are used for Mn$_3$GaN and Mn$_3$NiN, respectively. These estimations are chosen to yield a lattice parameter that approximately matches the experimental values in the paramagnetic state~\cite{LB1981}. We have found that such an effect enhances the transition temperature of Mn$_3$GaN up to 315K. This value is now higher than the transition temperature obtained for Mn$_3$NiN, which correctly captures the experimental trend ($T_{\text{tr},\text{Mn}_3\text{GaN}}>T_{\text{tr},\text{Mn}_3\text{NiN}}$, see Table \ref{Table3}). We conclude that accounting for the lattice thermal expansion is crucial for Mn$_3$GaN owing to its very large magnetovolume coupling.

%%%%%%%%%%%%%%%%%%%%%%%%%%%%%%%%%%%%%%%%%%%%%%%%%%%%%%%%%%%%%%%%%%%%
{\renewcommand{\arraystretch}{1.7}
\begin{table}[t]
    \centering
    \begin{tabular}{|cc|cc|}
    \hline
               &
               & $T_\text{tr}$ (K)
               & $\Delta\omega$ ($\%$)
               \\  \hline\hline
    Mn$_3$GaN & Experiment~\cite{PhysRevX.8.041035} & 290 & 1  \\
    & \thead{{\small Theory} \\  \textit{without lattice thermal expansion}} & 170 & 4  \\
    & \thead{{\small Theory} \\  \textit{with lattice thermal expansion}} & 315 & 3   \\
    \hline
    Mn$_3$NiN & Experiment~\cite{Matsunami1} & 262 & 0.4  \\ 
    & \thead{{\small Theory} \\  \textit{without lattice thermal expansion}} & 282 & 0.35  \\
    & \thead{{\small Theory} \\  \textit{with lattice thermal expansion}} & 290 & 0.35  \\
    \hline
    \end{tabular}
    \caption{The transition temperature and spontaneous volume change calculated at the first-order transition of Mn$_3$GaN and Mn$_3$NiN and their comparison with experiment.}
    \label{Table3}
\end{table}
}
%%%%%%%%%%%%%%%%%%%%%%%%%%%%%%%%%%%%%%%%%%%%%%%%%%%%%%%%%%%%%%%%%%%%

A major part of our results above agree qualitatively, and sometimes quantitatively, with experiment. However, our theory seems to substantially overestimate $U^{(4)}$ in Mn$_3$GaN~\cite{PhysRevX.8.041035}. Several factors can be behind this significant discrepancy. It is well known that Mn$_3$AN materials usually present a deficiency of nitrogen occupation~\cite{LB1981,doi:10.1088/1468-6996/15/1/015009} that potentially impacts the magnetic properties. Furthermore, minor chemical or positional disorder at the Mn sites can greatly modify the geometrical frustration of the magnetic interactions. We also expect, therefore, that a magneto-phonon coupling is another important component to accurately describe the magnetism of Mn$_3$AN. All these are aspects that we have not included in our calculations and that could explain the disagreement. Nevertheless, our approach distinguishes between electronic and magnetovolume mechanisms, describes how these depend on chemical composition, and correctly captures a major part of complicated features, such as the first-order character and the NVE.

As a final remark, we highlight that we have described the dependence of the internal magnetic energy in Eq.\ (\ref{EQ_EintMn3AN}) by considering the lowest possible orders of the expansion coefficients. However, we observed that such a dependence can be also described numerically well by taking a sixth order coefficient, $U^{(6)}$, instead of the fourth order, $U^{(4)}$. Considering $U^{(6)}$ provides the same qualitative and quantitative behavior for the paramagnetic-triangular antiferromagnetic phase transition in Mn$_3$GaN, with only tiny numerical differences. On the other hand, for Mn$_3$NiN the transition becomes continuous and falls right below a tricritical point separating first- and second- order behaviors, i.e.\ it becomes nearly discontinuous.

%%%%%%%%%%%%%%%%%%%%%%%%%%%%%%%%%%%%%%%%%%%%%%%%%%%%%%%%%%%
%%%%%%%%%%%%%%%%%%%%%%%%%%%%%%%%%%%%%%%%%%%%%%%%%%%%%%%%%%%
\section{Conclusions and outlook}
\label{Conc}
%%%%%%%%%%%%%%%%%%%%%%%%%%%%%%%%%%%%%%%%%%%%%%%%%%%%%%%%%%%
%%%%%%%%%%%%%%%%%%%%%%%%%%%%%%%%%%%%%%%%%%%%%%%%%%%%%%%%%%%

The development of new \textit{ab initio} theories describing the temperature evolution of magnetic materials is a major challenge greatly demanded by and strongly impacting the research of solid-state magnetism. It broadly includes the investigation of functional magnetic phase transitions between complex magnetic structures that are exploited in a wide range of technological applications.  
In this work we address this challenge by developing an approach that provides the \textit{ab initio} magnetic Gibbs free energy of a material from magnetically constrained supercell calculations. Its basis is the description of the statistical mechanics of local magnetic moments, assumed to evolve very slowly following the disordered local moment (DLM) picture~\cite{0305-4608-15-6-018}. We compute the internal magnetic energy of the material by performing averages of a density functional theory-based first-principles magnetic energy over a large but affordable number of noncollinear local moment configurations.

We have applied our supercell approach to study the ferromagnetic state of bcc iron and the triangular antiferromagnetic state present in the geometrically frustrated antiperovskite systems Mn$_3$AN (A = Ga, Ni). Our results are in good qualitative, and sometimes quantitative, agreement with experiment. Most importantly, we describe correctly the character of the magnetic phase transitions from the paramagnetic state, either continuous or discontinuous, and quantify its origin in terms of purely electronic and/or magnetostructural sources. We have found that the mechanism giving rise to the first-order character of Mn$_3$NiN arises purely from multisite magnetic interactions while this effect as well as a magnetovolume coupling play a major role in Mn$_3$GaN. Potential explanations for disagreements with experiment have been discussed.

Magnetically constrained supercell calculations are the principal computational component of our approach. This enables the application of our DLM theory using density functional theory codes based on a plane-wave basis (VASP in this work), i.e.\ beyond the Korringa-Kohn-Rostoker formalism and the coherent potential approximation.
Our approach is computationally expensive but it is already affordable by existing supercomputers.
Furthermore, satisfactory qualitative results are obtained using relatively small supercells.

An important advantage of
this DLM theory
is that the trial, mean-field, Hamiltonian prescribing the local moment averages [see Eq.\ (\ref{Eq_HIntro})] can be naturally extended beyond the simplest single-site Weiss field parameters~\cite{0305-4608-15-6-018}. In other words, magnetically constrained supercell calculations can be directly used to account for nonlocal magnetic correlations akin to the nonlocal coherent potential approximation~\cite{PhysRevB.73.205109,PhysRevB.94.224205,Staunton_2014}. Along these lines, spin-cluster expansions, which also follow an adiabatic approximation for the local moments, can be used to efficiently construct complex magnetic Hamiltonians~\cite{PhysRevB.69.104404}. Furthermore, magnetically constrained calculations directly output the magnetic torques added to sustain the transient orientational magnetic configurations. Our approach, therefore, also offers opportunities to be combined with spin-dynamics and related finite-temperature methods~\cite{PhysRevB.100.014105,PhysRevMaterials.5.053804}.
Finally, additional averages of the internal DFT energy can be performed over other degrees of freedom, such as the local moment magnitudes and the atom vibrations~\cite{PhysRevB.91.165132,PhysRevB.95.125109,Patrick_2014}. Pertinent time-scale separations for the lattice dynamics could then be considered~\cite{PhysRevLett.121.125902,PhysRevB.85.144404} to account for a magneto-phonon coupling~\cite{PhysRevLett.113.165503}.

%%%%%%%%%%%%%%%%%%%%%%%%%%%%%%%%%%%%%%%%%%%%%%%%%%%%%%%%%%%
%%%%%%%%%%%%%%%%%%%%%%%%%%%%%%%%%%%%%%%%%%%%%%%%%%%%%%%%%%%
\begin{acknowledgements}
We gratefully acknowledge helpful discussions with J.\ B.\ Staunton.
The authors acknowledge
funding from the ANR-DFG MAGIKID Project (Grant No.\ HI 1300/13-1)
and the computing time granted by the supercomputer of the Department of Computational Materials Design, operated by the Max Planck Computing and Data Facility in Garching.
\end{acknowledgements}
%%%%%%%%%%%%%%%%%%%%%%%%%%%%%%%%%%%%%%%%%%%%%%%%%%%%%%%%%%%
%%%%%%%%%%%%%%%%%%%%%%%%%%%%%%%%%%%%%%%%%%%%%%%%%%%%%%%%%%%

\appendix

%%%%%%%%%%%%%%%%%%%%%%%%%%%%%%%%%%%%%%%%%%%%%%%%%%%%%%%%%%%
%%%%%%%%%%%%%%%%%%%%%%%%%%%%%%%%%%%%%%%%%%%%%%%%%%%%%%%%%%%
\section{Calculations for smaller supercells}
\label{App1}

\begin{figure}[t]
\centering
\includegraphics[clip,scale=0.58]{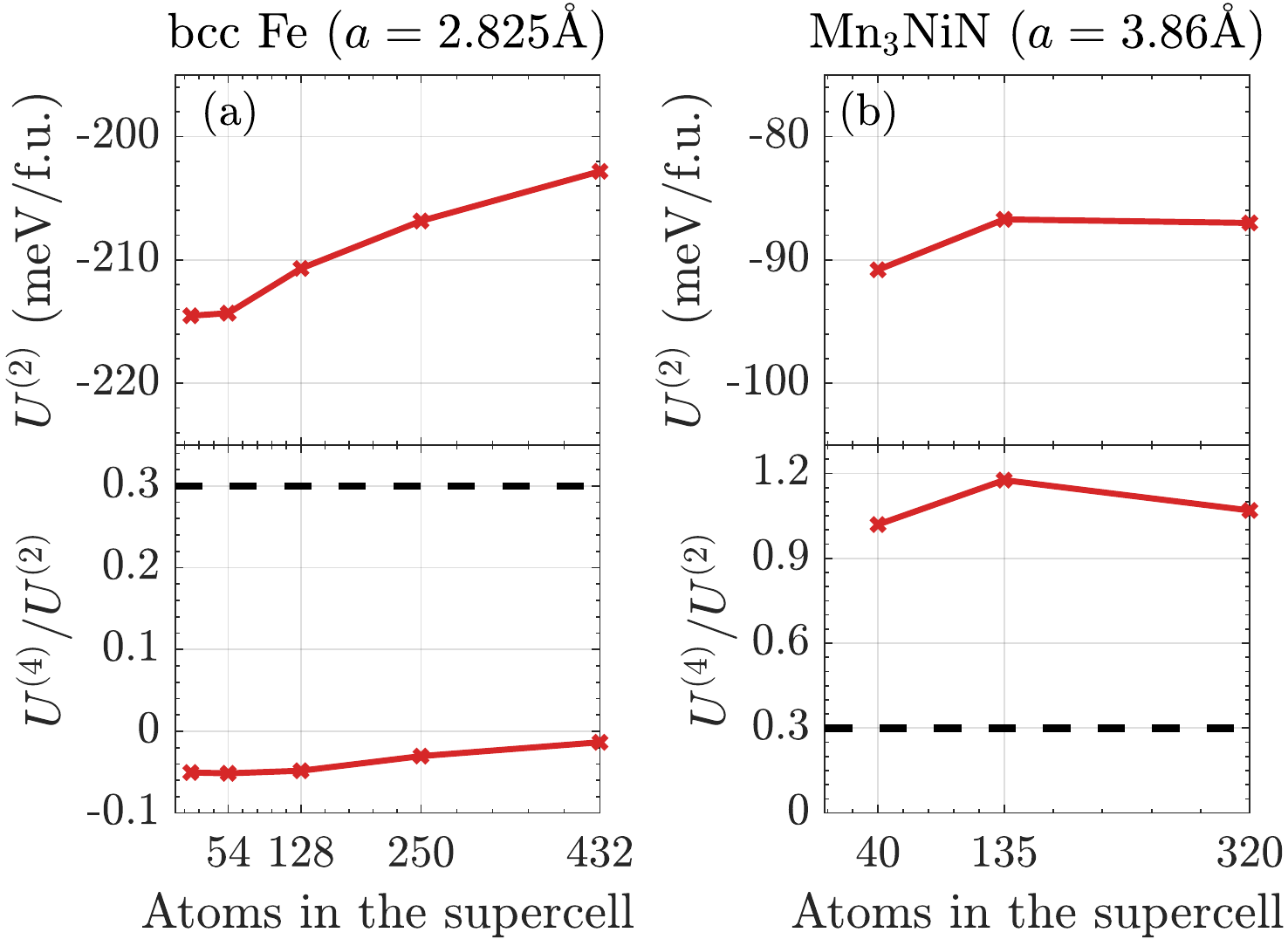}
\caption{
Dependence on the supercell size of the second and fourth order internal energy coefficients, $\{U^{(2)},U^{(4)}\}$, calculated for the ferromagnetic state of bcc Fe and the triangular antiferromagnetic state of Mn$_3$NiN. The lattice parameters are $a=2.825$\AA$\,$ and $a=3.86$\AA, respectively. The critical line (dashed) separating regions between second- ($\frac{U^{(4)}}{U^{(2)}}\leq\frac{3}{10}$) and first- ($\frac{U^{(4)}}{U^{(2)}}>\frac{3}{10}$) order magnetic phase transitions from the paramagnetic state is indicated in the lower panels.
}%
\label{FigApp1}
\end{figure}

Fig.\ \ref{FigApp1} shows the second and fourth order internal energy coefficients computed for ferromagnetic bcc Fe and for the triangular antiferromagnetic state of Mn$_3$NiN with smaller supercells. These calculations have been carried out for lattice parameters equal to $a=2.825$\AA$\,$ and $a=3.86$\AA, respectively. In Table \ref{TableApp} we provide the number of supercell snapshots used to carry out the average of the magnetic energy for every value of the magnetic order parameter. This number becomes larger for smaller supercells if similar statistical accuracy for the energy average is required. The Table also shows the number of $k$-points used within the Monkhorst-Pack grid sampling.

The same qualitative results are obtained even after reducing the supercell size to contain only a few tenths of atoms. For example, the ratio $U^{(4)}/U^{(2)}$ remains approximately zero and close to 1 for bcc Fe and Mn$_3$NiN, respectively. These values are below and above the critical condition $U^{(4)}/U^{(2)}=\frac{3}{10}$, which directly implies that the corresponding character of the magnetic phase transition from the paramagnetic state is second-order and first-order, respectively [see Eq.\ (\ref{EQ_cFOT2}) and Fig.\ \ref{FigFOTSOC}].
On the other hand, we observe a non-negligible quantitative change of $U^{(2)}$ for bcc Fe, i.e.\ there is a dependence of the computed transition temperature, $T_\text{tr}$, on the size of the supercell [see Eq.\ (\ref{EQ_TSOT})]. 
In this case, $T_\text{tr}$ decreases when the supercell becomes larger.
We conclude that qualitatively correct results can be already obtained for relatively small supercells, but that very accurate quantitative calculations require a large number of atoms. However, careful reassessment in this regard should be made for the evaluation of other magnetic materials.

%%%%%%%%%%%%%%%%%%%%%%%FIGURE%%%%%%%%%%%%%%%%%%%%%%%%%%%%%%%%%%%%%
%%%%%%%%%%%%%%%%%%%%%%%%%%%%%%%%%%%%%%%%%%%%%%%%%%%%%%%%%%%%%%%%%%%%

{\renewcommand{\arraystretch}{1.7}
\begin{table}[b]
    \centering
    \begin{tabular}{|c|ccc|}
    \hline
      Material & \thead{Atoms in the \\ supercell} & \thead{Number of \\ snapshots, $N_\text{MC}$} & $k$-mesh  \\
    \hline
      bcc Fe & 16 & 50 & $6\times 6\times 6$ \\ 
        & 54 & 30 & $4\times 4\times 4$ \\ 
        & 128 & 30 & $3\times 3\times 3$ \\
        & 250 & 25 & $2\times 2\times 2$ \\
        & 432 & 10 & $2\times 2\times 2$ \\
    \hline
      Mn$_3$NiN & 40 & 80 & $3\times 3\times 3$ \\
       & 135 & 25 & $2\times 2\times 2$ \\
       & 320 & 10 & $2\times 2\times 2$ \\
    \hline
    \end{tabular}
    \caption{
Number of supercell snapshots and $k$-mesh used to carry out the average of the magnetic energy for the different choices of supercell sizes in Fig.\ \ref{FigApp1}. The higher the number of atoms in the supercell, the higher the number of necessary snapshots to achieve satisfactory accuracy for the average.}
    \label{TableApp}
\end{table}
}

\newpage

\bibliography{./bibliography.bib}

\end{document}